\documentclass[preprint,5p,times,twocolumn,authoryear]{elsarticle}

\usepackage[export]{adjustbox}

\usepackage{amssymb,amsmath}

\usepackage{bm}

\usepackage{cancel}

\usepackage[dvipsnames]{xcolor}

\usepackage[labelfont=bf]{caption}

\usepackage[
    authormarkuptext=name,
    authormarkupposition=left,
    commentmarkup=uwave,
    defaultcolor=BrickRed,
    final
]{changes}

\usepackage[colorlinks]{hyperref}
\usepackage[nameinlink,capitalise]{cleveref}
\usepackage{xurl}

\usepackage[
    procnumbered, %% YAVUZ: It seemed to me that we should not restart counting algorithms when moving on from Background to Method.
    ruled,
    lined,
    linesnumbered,
    inoutnumbered, %% YAVUZ: IMHO, this does not make any sense, but still the case in recent papers.
    commentsnumbered,
    longend %% YAVUZ: IMHO, this does not make any sense, but still the case in recent papers.
]{algorithm2e}

\usepackage[byauthor]{credits}

\usepackage[switch]{lineno}

\usepackage{orcidlink}

\usepackage[aboveskip=2pt,belowskip=-2pt]{subcaption}

\usepackage{pbalance}

\usepackage{csquotes}

\usepackage{tcolorbox}

\usepackage{siunitx}

\usepackage{calc}

\usepackage{hyphenat}

\usepackage{booktabs}

\usepackage{colortbl}

\usepackage[T1]{fontenc}

\usepackage{multirow}

\definecolor{GraphFrameBG}{RGB}{254,250,240}

\newenvironment{qblock}{%
\begin{displayquote}\begin{tcolorbox}[%
    width=\linewidth,% Take the width as parameter
    size=small,% Small inside padding
    colframe=Black,% Text color
    colback=GraphFrameBG,% Fill color
    nofloat,% tcolorbox is not a float because we put the tcolorbox inside another float.
    fontupper=\itshape
]}{\end{tcolorbox}\end{displayquote}}

\Crefname{algocf}{Algorithm}{Algorithms}
\Crefname{assumption}{Assumption}{Assumptions}

\newdefinition{definition}{Definition}
\newdefinition{assumption}{Assumption}
\newdefinition{theorem}{Theorem}

\newenvironment{elsfigure}[1][t]{\begin{figure}[#1]\centering}{\end{figure}}
\newenvironment{elsfigure*}[1][t]{\begin{figure*}[#1]\centering}{\end{figure*}}
\newenvironment{elstable}[1][t]{\begin{table}[#1]\centering}{\end{table}}
\newenvironment{elstable*}[1][t]{\begin{table*}[#1]\centering}{\end{table*}}

\newenvironment{graphframe}[1][\linewidth]{%
\begin{tcolorbox}[%
    width=#1,% Take the width as parameter
    size=small,% Small inside padding
    colframe=Black,% Text color
    colback=GraphFrameBG,% Fill color
    center,% Try to put the tcolorbox at the center
    halign=center,% Try the put content at tcolorbox's center
    nofloat,% tcolorbox is not a float because we put the tcolorbox inside another float.
    before skip=0pt,% Do not put any outside padding before tcolorbox
    after skip balanced=0pt% Do not put additional outside padding, but respect the paper's padding
]}{\end{tcolorbox}}

\definechangesauthor[name={Onur}, color=BrickRed]{Onur}
\definechangesauthor[name={Yavuz}, color=Orange]{Yavuz}
\definechangesauthor[name={Mutlu}, color=ProcessBlue]{Mutlu}
\definechangesauthor[name={Serge}, color=RoyalPurple]{Serge}
\definechangesauthor[name={Franz}, color=Emerald]{Franz}
\definechangesauthor[name={Issue {\#}1}]{I1}
\definechangesauthor[name={Issue {\#}2}]{I2}
\definechangesauthor[name={Issue {\#}3}]{I3}
\definechangesauthor[name={Issue {\#}4}]{I4}
\definechangesauthor[name={Issue {\#}5}]{I5}
\definechangesauthor[name={Issue {\#}6}]{I6}
\definechangesauthor[name={Issue {\#}7}]{I7}
\definechangesauthor[name={Issue {\#}8}]{I8}
\definechangesauthor[name={Issue {\#}9}]{I9}
\definechangesauthor[name={Issue {\#}10}]{I10}
\definechangesauthor[name={Issue {\#}11}]{I11}
\definechangesauthor[name={Issue {\#}12}]{I12}
\definechangesauthor[name={Issue {\#}13}]{I13}
\definechangesauthor[name={Issue {\#}14}]{I14}
\definechangesauthor[name={Issue {\#}15}]{I15}
\definechangesauthor[name={Issue {\#}16}]{I16}
\definechangesauthor[name={Issue {\#}17}]{I17}
\definechangesauthor[name={Issue {\#}18}]{I18}
\definechangesauthor[name={Issue {\#}19}]{I19}
\definechangesauthor[name={Issue {\#}20}]{I20}
\definechangesauthor[name={Issue {\#}21}]{I21}
\definechangesauthor[name={Issue {\#}22}]{I22}
\definechangesauthor[name={Issue {\#}23}]{I23}
\definechangesauthor[name={Issue {\#}24}]{I24}
\definechangesauthor[name={Issue {\#}25}]{I25}
\definechangesauthor[name={Issue {\#}26}]{I26}
\definechangesauthor[name={Issue {\#}27}]{I27}
\definechangesauthor[name={Issue {\#}28}]{I28}
\definechangesauthor[name={Issue {\#}29}]{I29}
\definechangesauthor[name={Issue {\#}30}]{I30}
\definechangesauthor[name={Issue {\#}31}]{I31}
\definechangesauthor[name={Issue {\#}32}]{I32}
\definechangesauthor[name={Issue {\#}33}]{I33}
\definechangesauthor[name={Issue {\#}34}]{I34}
\definechangesauthor[name={Issue {\#}35}]{I35}
\definechangesauthor[name={Issue {\#}36}]{I36}
\definechangesauthor[name={Issue {\#}37}]{I37}
\definechangesauthor[name={Issue {\#}38}]{I38}
\definechangesauthor[name={Issue {\#}39}]{I39}
\definechangesauthor[name={Issue {\#}40}]{I40}
\definechangesauthor[name={Issue {\#}41}]{I41}
\definechangesauthor[name={Issue {\#}42}]{I42}
\definechangesauthor[name={Issue {\#}43}]{I43}
\definechangesauthor[name={Issue {\#}44}]{I44}
\definechangesauthor[name={Issue {\#}45}]{I45}
\definechangesauthor[name={Issue {\#}46}]{I46}
\definechangesauthor[name={Issue {\#}47}]{I47}
\definechangesauthor[name={Issue {\#}48}]{I48}
\definechangesauthor[name={Issue {\#}49}]{I49}
\definechangesauthor[name={Issue {\#}50}]{I50}
\definechangesauthor[name={Issue {\#}51}]{I51}
\definechangesauthor[name={Issue {\#}52}]{I52}
\definechangesauthor[name={Issue {\#}53}]{I53}
\definechangesauthor[name={Issue {\#}54}]{I54}
\definechangesauthor[name={Issue {\#}55}]{I55}
\definechangesauthor[name={Issue {\#}56}]{I56}
\definechangesauthor[name={Issue {\#}57}]{I57}
\definechangesauthor[name={Issue {\#}58}]{I58}
\definechangesauthor[name={Issue {\#}59}]{I59}
\definechangesauthor[name={Issue {\#}60}]{I60}
\definechangesauthor[name={Issue {\#}61}]{I61}
\definechangesauthor[name={Issue {\#}62}]{I62}
\definechangesauthor[name={Issue {\#}63}]{I63}
\setanonymousname{}

\SetAlFnt{\small\rmfamily}
\newenvironment{elsalgorithm}[1][t]{%[htbp]
\begin{algorithm}[#1]%
    \DontPrintSemicolon% Dont end every statement with a ";"
    \SetAlgoCaptionSeparator{ }% Put a space between the caption content and the caption label
    \SetAlgoNlRelativeSize{0}% Set line number size as the body font size 
    \SetNlSty{}{}{:}% Put a ":" after every line number
    \SetInd{.75em}{.75em}% Indentation size
    \SetKw{And}{and}%
    \SetKw{Whl}{while}%
    \SetKw{St}{s.t.}%
    \SetKw{To}{to}%
    \SetKw{nForEach}{foreach}% One liner for
    \SetKw{nDo}{do}% One liner do
    \SetKwFor{nRepeat}{repeat}{times}{end repeat}% repeat N times --- end repeat
}{\end{algorithm}}

\DeclareMathSymbol{\shortminus}{\mathbin}{AMSa}{"39}

\newcommand{\tsuper}[1]{\raise0.6ex\hbox{\scriptsize #1}}

\newcommand{\N}{\mathbb{N}}

\newcommand{\symsetof}{:}
\newcommand{\setof}[2]{\lbrace #1 \symsetof #2 \rbrace}

\newcommand{\ltrl}[1]{\mathit{#1}}

\newcommand{\func}[1]{\mathrm{#1}}

\newcommand{\tgt}[1]{\func{tgt}(#1)}

\newcommand{\pre}[1]{\func{pre}(#1)}

\newcommand{\len}[1]{|#1|}

\newcommand{\symconcat}{\bm\cdot}
\newcommand{\concat}[2]{#1 \symconcat #2}

\newcommand{\symcycle}{\bm\bigcirc}
\newcommand{\cycle}[1]{\mathop{\symcycle}\!#1}
\newcommand{\ncycle}[1]{\mathop{\cancel{\symcycle}}\!#1}

\newcommand{\symrot}{\bm\circlearrowleft}
\newcommand{\rot}[1]{\mathop{\symrot}#1}

\newcommand{\simplegraph}[3]{#1 = (#2, #3), #3 \subseteq #2^2}

\newcommand{\multigraph}[4]{#1 = (#2, #3, #4), #4 \subseteq #2 \times #3 \times #2}

\newcommand{\graphmodel}[5]{#1 = (#2, #5), \simplegraph#2#3#4, #5 \in #3}

\newcommand{\symsubpath}{\sqsubseteq}

\newcommand{\subpath}[2]{#1 \symsubpath #2}

\newcommand{\symsplice}{\bm\odot}
\newcommand{\splice}[2]{#1 \symsplice #2}
\newcommand{\spliceG}[3]{#1 \symsplice_{#3} #2}

\newcommand{\crit}[1]{\ltrl{TR}_{\text{#1}}}

\newcommand{\hypergraph}[4]{#1 = (#3, #4), #3 \subseteq P(#2), #4 \subseteq #3^2}

\newcommand{\rf}{\func{RF}}
\newcommand{\rr}{\func{RR}}
\newcommand{\cov}[1]{\func{C_{#1}}}
\newcommand{\dlen}[1]{\len{\len{#1}}}

\newcommand*{\defeq}{\stackrel{\mathsmaller{\mathsf{def}}}{=}}

\newcommand*{\defiff}{\stackrel{\mathsmaller{\mathsf{def}}}{\leftrightarrow}}

\journal{Journal of Systems and Software}
            
\begin{document}

\onecolumn
\begin{frontmatter}

\hypersetup{pdftitle={DJPlus: Generating minimal test suites for strong coverage criterion in graph models}}
% \title{DJPlus: Generating minimal test suites for \replaced[id={I5}]{the edge-pair criterion}{strong coverage criteria} in graph models}
\title{DJPlus: Generating minimal test suites for strong coverage criteria in graph models}

\hypersetup{pdfauthor={Yavuz K{\"o}ro\u{g}lu, Mutlu Beyaz{\i}t, Onur K{\i}l{\i}n\c{c}\c{c}eker, Serge Demeyer, Franz Wotawa}}

\author[itu]{Yavuz K{\"o}ro\u{g}lu\corref{cor2}\orcidlink{0000-0001-9376-0698}}\ead{yakoroglu@itu.edu.tr}%
\author[antwerp]{Mutlu Beyaz{\i}t\corref{cor1}\orcidlink{0000-0003-2714-8155}}\ead{mutlu.beyazit@uantwerpen.be}%
\author[antwerp]{Onur K{\i}l{\i}n\c{c}\c{c}eker\corref{cor1}\orcidlink{0000-0001-5996-4398}}\ead{onur.kilincceker@uantwerpen.be}%
\author[antwerp]{Serge Demeyer\corref{cor1}\orcidlink{0000-0002-4463-2945}}\ead{serge.demeyer@uantwerpen.be}%
\author[tugraz]{Franz Wotawa\corref{cor1}\orcidlink{0000-0002-0462-2283}}\ead{wotawa@tugraz.at}%

\affiliation[itu]{
    organization={Department of Computer Engineering, {\.I}stanbul Technical University},
    addressline={Ayaza\u{g}a},
    city={Maslak},
    postcode={34469},
    state={{\.I}stanbul},
    country={T{\"u}rkiye}
}

\affiliation[antwerp]{
    organization={University of Antwerp and Flanders Make vzw},
    addressline={Middelheimlaan 1},
    city={Antwerpen},
    postcode={2020},
    %state={},
    country={Belgium}
}

\affiliation[tugraz]{
    organization={Institute of Software Engineering and Artificial Intelligence, Graz University of Technology},
    addressline={Inffeldgasse 16b/2},
    city={Graz},
    postcode={8010},
    state={Steiermark},
    country={Austria}
}

\cortext[cor1]{Corresponding author.}
\cortext[cor2]{Principal corresponding author.}

\begin{abstract}%
Automated test generation from graph models is essential to model-based testing. 
In this type of testing, graph coverage ensures test suite strength but also results in long test cases that take time to execute on the system under test.
We propose a novel optimization-driven method, DJPlus, which generates reduced test suites while satisfying given graph-based test requirements.
We implement DJPlus and show the feasibility of edge-pair criterion, a stronger coverage criterion than vertex or edge criteria, on four realistic systems, while prime path criterion poses scalability issues. 
Our evaluation reveals that the alternative methods generate 2 to 26 times more redundant test steps than DJPlus and DJPlus decreases test execution times by reducing the number of test steps\replaced[id={I11}]{. These results show that DJPlus is a positive step towards}{, indicating significant benefits in} tackling the challenges of model-based testing at an industrial scale.%
\end{abstract}

\begin{graphicalabstract}%
\includegraphics[frame,width=\textwidth,height=.4\textwidth]{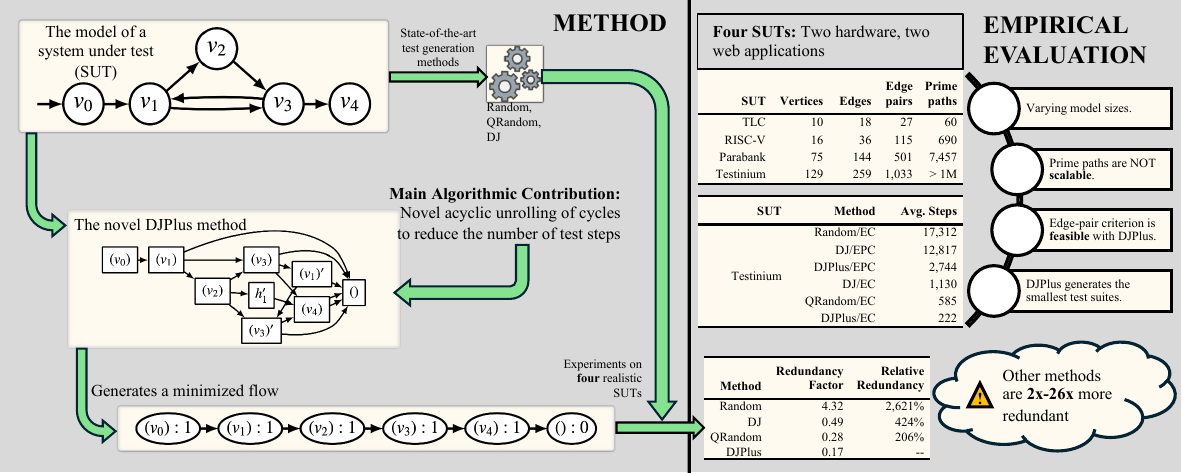}%
\end{graphicalabstract}

\begin{highlights}%
\item The novel DJPlus method generates minimal number of test cases for strong coverage.
\item Alternative methods generate 2 to 26 times more redundant test steps.
\item \replaced[id={I10}]{The e}{E}dge-pair criterion is shown to be feasible for realistic SUTs using DJPlus.
\item \replaced[id={I10}]{The p}{P}rime path criterion is shown to pose significant scalability issues in experiments.
\item DJPlus decreases test execution times by reducing the number of test steps.
\end{highlights}


\end{frontmatter}

%\linenumbers

\section{Introduction}\label{sec:introduction}
Model\hyp{}based testing (MBT) is a software testing approach that uses models to generate test cases for a system under test (SUT). 
As a research subject, it is more than $50$ years old~\citep{Elmendorf1970:IBM}.
Benefits of MBT include a high degree of test automation, high-quality tests, and reduced manual testing effort~\citep{Dalal+1999:ICSE}. \replaced[id={I31}]{However, the complexity of MBT approaches allow industrial adoption only in small steps~\citep{Janicki+2012:STVR}. Test cases quickly explode especially in coverage directed testing~\citep{Mlynarski+2012:AdvC}. So, there is a tradeoff between MBT complexity and its scalability~\citep{Utting+2016:AdvC}.}{Despite these benefits, one study reports that MBT is still not a mainstream testing approach (Neto et al., 2008), while another claims that the potential benefits of MBT do not generalize to all its applications (Vaandrager, 2006). Then, clearly, there is much to improve in MBT.}

Although \replaced[id={I32}]{a model of an SUT (SUT model)}{an SUT model} can be any abstract representation that captures some system behavior, a graph\added[id={I33}]{ (a transition system)} is a typical model in MBT~\citep{Alegroth+2022:ESE}, where a path in the graph constitutes a test case and the elements of this path are test steps.
A test generation method produces a set of test cases, i.e., a test suite, from the graph model.
One way to assess the strength of a test suite is coverage\added{-based criteria}~\citep{MasriZaraket:2016:ElsevierAdvC}.
While many MBT methods use \added[id={I10}]{the }vertex or edge \deleted{coverage }criteria~\citep{Koroglu+2025:ICSTW:AMOST}, the literature also features \added{the }edge\hyp{}pair and prime path criteria, which ensure stronger tests~\citep[Part 2, Ch.~7]{AmmanOffut2008:CUP}.
We call a \deleted{coverage }criterion stronger than another if achieving the former guarantees achieving the latter. 
MBT methods often resort to relatively weaker criteria \deleted{like edge coverage }due to scalability concerns caused by the execution times of the generated test suite.
Still, according to an experience report, the time to \replaced[id={I27}]{satisfy the}{achieve $100\%$} edge \replaced{criterion}{coverage} via fully\hyp{}automated test generation and execution on a web application was six hours~\citep{Garousi+2021:JSS}, revealing some scalability issues even under weak criteria. \replaced{This result aligns well with the conclusions of another study}{Another study confirms} that MBT practice leads to large test suites even for weak criteria and proposes test selection techniques to minimize the test suite after the generation phase~\citep{Hemmati+2010:FSE}.

In this study, we focus on a method to generate a \deleted[id={I41}]{minimal} test suite for a graph coverage criterion \added[id={I41}]{whose total number of test cases are minimized first and then its test steps}~\citep{Li+:2012:ICST}. 
The Dwarakanath and Jankiti (DJ) method is the state\hyp{}of\hyp{}the\hyp{}art approach that generates a minimal number of test cases from a graph model~\citep{DwarakanathJankiti2014:ICTSS}\replaced{, but}{.
But,} it does not optimize the number of test steps within the test cases, leaving room for improvement in terms of scalability.

To the best of our knowledge, popular MBT approaches do not yet utilize the DJ method.
GraphWalker\footnote{See~\url{https://graphwalker.github.io}.} is a typical example of an open\hyp{}source graph\hyp{}ba\-sed MBT tool, which provides the functionality to create graph models, generate abstract test cases from these graphs, and execute them on the SUT. We focus on GraphWalker throughout this study because the literature indicates widespread use of GraphWalker in many domains, including electronic circuits~\citep{Darwish+2017:ICST}, command\hyp{}line tools~\citep{Castro-Cabrera+2022:ICSEA}, mobile applications~\citep{Gudmundsson+2016:PrePostIFM}, and safety\hyp{}critical systems~\citep{Zafar+2023:ENASE}. GraphWalker comes with two built\hyp{}in random test generation methods, Random and QRandom (a variant of Random utilizing shortest paths), and does not implement the DJ method.

In light of all the above discussion, we present the following contributions to the literature.

\begin{enumerate}
    \item \added[id={I2}]{\emph{Algorithmic novelty:}} We present DJPlus, \replaced{an improvement to}{a novel method that generates a minimal number of test cases with a reduced amount of test steps. So, DJPlus improves} the state\hyp{}of\hyp{}the\hyp{}art DJ method\deleted{, which minimizes only the number of test cases}. \added{DJPlus, like DJ, generates the minimum number of test cases but also reduces the number of test steps.} We explain DJPlus in detail, with examples.
    \item \added[id={I3}]{\emph{GraphWalker integration:}} We implement the state-of-the-art DJ and our novel DJPlus methods to generate test cases for the popular GraphWalker environment. We name this implementation GWPlus, which replaces GraphWalker's built\hyp{}in random test generation. With this implementation, we bridge the gap between the most recent theoretical developments and a practical MBT tool.
    \item \deleted[id={I3}]{We evaluate known graph coverage criteria and demonstrate that the prime path criterion does not scale well as the SUT model grows. However, the edge\hyp{}pair criterion, a stronger criterion than the vertex and edge criteria, is still feasible for our experimental SUT models. 4. }\added{\emph{Empirical evaluation:}} We perform experiments on four realistic SUTs: two web applications (Parabank and Testinium) and two hardware applications (TLC and RISC-V), and show that DJPlus achieves the same coverage with \replaced[]{fewer}{the fewest} steps. We also show that other methods are 2 to 26 times more redundant than DJPlus, while DJPlus 
    decreases the test execution times by reducing the number of test steps.
\end{enumerate}

\added[id={I52}]{Along with our contributions to the literature, we answer the following research questions (RQs):}

\deleted[id={I53}]{\textbf{RQ1:} (Scalability of coverage criteria) How do vertex, edge, edge\hyp{}pair, and prime path criteria scale as the model size grows?}

\deleted{Motivation. In this study, we consider edge\hyp{}pair and prime path criteria as stronger alternatives to vertex and edge criteria. However, as we described in Section II.A, strong coverage comes with a tradeoff in additional test steps. This tradeoff is significant because, as we discussed in Section 1, some studies resort to weak criteria due to concerns about scalability (Garousi et al. 2021, Garousi et al., 2010). In short, with this research question, we aim to strike a balance between scalability and coverage, seeking a strong coverage criterion that is also feasible for large graph models.}

\deleted{Method. We compare the number of test requirements generated by GWPlus for four realistic SUT models with varying sizes.
We sort these models from smallest to largest, and show the trends in the number of test requirements for the four coverage criteria in the research question.}

\begin{qblock}
    \textbf{RQ\replaced{1}{2}:} (Feasibility\deleted{ of strong criteria}) \replaced{Is}{Are} \added{the} edge\hyp{}pair \deleted{and prime path }\replaced{criterion}{criteria} feasible for realistic systems?
\end{qblock}

\added{With this RQ, we investigate the costs associated with targeting the edge-pair criterion instead of the vertex or the edge criteria. To answer this RQ, we first compare the number of test requirements as the model grows under different criteria. Then, we compare the number of test steps generated by the established state-of-the-art, GraphWalker, for the edge criterion and by our proposed method, DJPlus, for the edge-pair criterion. We argue that any number of test steps comparable to the established state-of-the-art should remain feasible.}

\deleted{Motivation. Even if a strong coverage criterion is scalable, it is theoretically costlier than a weaker one. With this research question, we investigate the costs associated with targeting edge\hyp{}pair and prime path criteria compared to a more widely used approach.}

\deleted{Method.
We compare the total number of test steps of tests generated for four realistic SUTs by DJPlus, targeting strong coverage, and GraphWalker, targeting edge coverage.}

\begin{qblock}
    \textbf{RQ\replaced{2}{3}:} (Improvement to the state-of-the-art) Does DJPlus generate test suites with fewer test steps than DJ?
\end{qblock}

\added{DJPlus does not always generate the minimal number of test steps to satisfy a criterion. So, we perform an empirical evaluation of Random, QRandom, DJ, and DJPlus to show that DJPlus generates fewer test steps compared to its alternatives.}

\deleted{Motivation. We demonstrate in Section 3 how DJPlus generates test suites with fewer test steps than DJ. However, that is not an indicator of the general minimality of DJPlus. With this research question, we aim to compare the performance of DJ and DJPlus on realistic systems.}

\deleted{Method. We compare test suites generated by DJ and DJPlus for four realistic SUTs, targeting $100\%$ vertex, edge, and edge\hyp{}pair coverage.}

\begin{qblock}
    \textbf{RQ\replaced{3}{4}:} (Redundancy) Are DJ, Random, and QRandom redundant compared to DJPlus, and by how much?
\end{qblock}

\added{Since none of the known test generation methods we investigate are optimal, including DJPlus, all of them generate test steps that are redundant in terms of their target criteria. }\added[id={I6}]{However, these test steps could still have some utility in terms of fault detection or satisfying a stronger coverage criterion. In this study, we define a redundancy measure called edge-pair redundancy and use it as an indicator of the redundancy of test steps that target the vertex and the edge criteria.}

\deleted{Motivation. With this research question, we aim to quantify the benefits of DJPlus compared to its alternatives.}

\deleted{Method. We compare the number of test steps and report the average relative redundancy of other methods over DJPlus. The rationale behind this measure is the idea of coverage\hyp{}based redundancy, i.e., test steps that re\hyp{}trigger what is already covered. We explain redundancy in Section IV.C and give further details about its computation in Appendix C.}

\begin{qblock}
    \textbf{RQ\replaced{4}{5}:} (Reduction of test execution times) Does DJPlus decrease test execution times?
\end{qblock}

\added[id={I58}]{DJPlus attempts to indirectly decrease test execution times by reducing the number of test steps. With this RQ, we explore whether reducing the number of test steps cause a decrease in test execution times, through an effect size analysis between our empirical results.}

\deleted{Motivation. Our approach does not consider the varying execution times of different test steps which potentially hinder our claim of decreasing execution times through reducing test steps. With this research question, we investigate the relation between the reduced test suites of DJPlus and the decreases in test execution times.}

\deleted{Method. We execute tests generated by GraphWalker's built\hyp{}in methods, DJ, and DJPlus, ten times each. Then, we perform a correlation analysis between the number of test steps and average test execution times to demonstrate the strong relation between minimizing test steps and reducing test execution times.}

We organize the rest of this paper starting with a discussion of the related work in \cref{sec:relatedwork}. In \cref{sec:background}, we explain the concepts necessary to understand our method. We describe our method, pose and address research questions, elaborate on the results, and conclude with a summary in \cref{sec:method,sec:evaluation,sec:discussion,sec:conclusions}.

\section{Related work}\label{sec:relatedwork}

We now motivate our focus on MBT with graph models, generating minimal number tests from graphs, graph coverage, and GraphWalker in the following sections.

\subsection{MBT with graph models}
\label{sec:relwork:mbt}

Graph models in MBT are as old as MBT itself~\citep{Elmendorf1970:IBM}. One systematic mapping study reports that most MBT tools use diagrams, charts, or finite state machines, which are all graph models~\citep{DiasNeto+2007:WEASELTech}. Some graph models, like unified model diagrams, have the added benefit of visualization and interpretability, facilitating software development~\citep{Rumpe2016:Springer}. Overall, graph models remain central to MBT and model-driven engineering.

\subsection{Minimal test generation from graphs}

\label{sec:relwork:minimal}
\begin{elstable*}
    \centering
    \caption{\added[id={I13}]{A comparison of minimal test generation approaches for graph models.}}
    \label{tab:relwork}
    %\resizebox{.8\linewidth}{!}{
    \begin{tabular}{r|cccc|cc|c}
         \multicolumn{1}{c|}{\multirow{2}{*}{\textbf{Method}}} & \multicolumn{4}{c|}{\textbf{Criterion}} & \multicolumn{2}{c|}{\textbf{Minimization}} &\\
         \cline{2-7}
         & \textbf{VC} & \textbf{EC} & \textbf{EPC} & \textbf{PPC} & \textbf{\# Test Cases} & \textbf{\# Test Steps} & \\
         \hline
         \citep{AhoLee:1987} & $\checkmark$ & & & & & $\checkmark$ & \\
         \citep[Part 2, Ch.~7]{AmmanOffut2008:CUP} & & & & $\checkmark$ & & $\checkmark$ \\
         DJ~\citep{DwarakanathJankiti2014:ICTSS} & $\checkmark$ & $\checkmark$ & $\checkmark$ & $\checkmark$ & $\checkmark$ & \\
         DJPlus (Our proposed approach) & $\checkmark$ & $\checkmark$ & $\checkmark$ & $\checkmark$ & $\checkmark$ & $\checkmark$ & \\
         \hline
         \multicolumn{8}{r}{\footnotesize \textbf{VC:} Vertex Criterion, \textbf{EC:} Edge Criterion, \textbf{EPC:} Edge-Pair Criterion, \textbf{PPC:} Prime Path Criterion}\\
    \end{tabular}
    %}
\end{elstable*}

Minimal test generation for a graph coverage criterion has a history comparable to MBT~\citep{NtafosHakimi1979:TSE}.
\replaced[id={I2}]{\cref{tab:relwork} provides a non-comprehensive comparison of previous minimal test generation approaches for graph models. According to this table, o}{O}ne of the earliest methods invented for this purpose targets only vertex criterion~\citep{AhoLee:1987}. 
Another method targets prime path criterion but does not attempt to minimize the test cases~\citep[Part 2, Ch.~7]{AmmanOffut2008:CUP}.
The latest development in this area is the DJ method, which generates the fewest test cases and targets any set of test requirements rather than a specific coverage criterion~\citep{DwarakanathJankiti2014:ICTSS}.
However, DJ-generated test cases still contain many redundant test steps.
To the best of our knowledge, ours is the first work to tackle both minimizing the number of test cases and reducing the number of test steps while satisfying any given set of test requirements. 

\subsection{Graph coverage}
\label{sec:relwork:cov}

The literature includes several coverage criteria for graph models and defines a subsumption relation among these depending on their test strengths~\citep[Part 2, Ch.~7]{AmmanOffut2008:CUP}.
\replaced[id={I10}]{The v}{V}ertex and edge criteria are weak in this relation, while \added[id={I10}]{the }edge-pair and prime path criteria are stronger.
\replaced[id={I10}]{The p}{P}rime path criterion has the added benefit of subsuming data-flow criterion and facilitating logic coverage~\citep{Kaminski+2010:SERP}.
To the best of our knowledge, ours is the first MBT approach targeting \added[id={I10}]{the }edge-pair and prime path criteria \replaced[id={I12}]{with a reduced number test steps and a minimized number of test cases}{in general MBT practice}.

\subsection{GraphWalker}
\label{sec:relwork:gw}

GraphWalker is a popular MBT tool for modeling or test generation within a diverse set of domains, including the testing of programmable logic controllers~\citep{Frey+2012:ETFA}, web applications~\citep{Schur+2013:FSE}, graphical user interfaces~\citep{Kilincceker+2021:IEEEAccess}, data access tools~\citep{Lindvall+2015:ICSE}, electronic control circuitry~\citep{Darwish+2017:ICST}, service orchestrations~\citep{Leal+2020:PRDC}, mobile applications~\citep{Karlsson+2021:ICSTW}, command-line tools~\citep{Castro-Cabrera+2022:ICSEA}, information and communications technology supply chains~\citep{Bicchierai+2023:ISSRE}, ground system software~\citep{Gudmundsson+2015:ISSE}, and autonomous vehicles~\citep{ChetouaneWotawa2023:ICST}.
One study performs model checking in a GraphWalker environment using UPPAAL~\citep{Tiwari+2022:APSEC}.

Further than its apparent popularity, one experience report \replaced[id={I35}]{found GraphWalker to be only one of two tools out of ten that features all the essential criteria for MBT, while the other candidate, TestOptimal, unlike GraphWalker, is not open-source and easily modifiable~\citep{Garousi+2021:JSS}}{indicates that GraphWalker is preferable in practice thanks to its simplicity and flexibility~(Garousi et al., 2021)}.
However, the same study also shows a case where the total test execution time of an SUT was six hours to \replaced[id={I27}]{satisfy the edge criterion with GraphWalker}{achieve $100\%$ edge coverage}, revealing potential scalability problems of test generation from graph models\added[id={I36}]{ as the model grows in size}.

Some studies focus on the costs of GraphWalker-generated test cases, where one study reports that they contain many more test steps than manually crafted test cases that reach the same coverage~\citep{Karlsson+2022:ACM}.
A more recent study demonstrates that GraphWalker-generated test cases contain many redundant test steps, i.e., test steps that do not increase vertex or edge coverage~\citep{Koroglu+2025:ICSTW:AMOST}.
This study investigates the possible benefits of these redundant test steps by measuring their contributions to edge\hyp{}pair coverage, reporting that the cost of an additional percent of edge\hyp{}pair coverage increases exponentially in terms of the number of test steps.
Thus, the study concludes that a deterministic, optimization-based test generation method is necessary to avoid the test execution costs while satisfying a strong graph coverage criterion.

\section{Background}\label{sec:background}

\replaced[id={I9}]{We now explain how the state-of-the-art DJ method generates the minimal number of test cases but how it fails to eliminate redundancies within these test cases, on an example graph. We provide more rigorous definitions to facilitate the understanding the DJ methodology in \ref{sec:graphs}.}{We now explain the concepts necessary to understand DJPlus. For more rigorous definitions of these concepts, we refer to Appendix A.}\replaced[id={I17}]{The concepts used in this section are aligned with the original DJ study~\cite{DwarakanathJankiti2014:ICTSS}. They also}{These concepts} closely follow the wider graph\hyp{}based testing theory established in the literature~\citep{AmmanOffut2008:CUP}.

\begin{elsfigure}
    \begin{graphframe}[.75\linewidth]
        \includegraphics{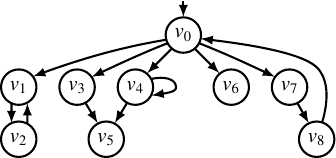}
    \end{graphframe}
    \caption{An example graph \replaced[id={I13}]{that models the behaviors of a system under test (SUT)}{model with nine vertices and twelve edges}.}\label{fig:examplegraph}
\end{elsfigure}

\cref{fig:examplegraph} depicts an example graph model we designed solely to facilitate our explanations. It features nine vertices representing nine different SUT behaviors, where $v_0$ is the entry vertex. Twelve edges connect these behaviors. Notice that every vertex is reachable from $v_0$, and $v_4$ has a loop. An example path of length four \added{of this graph} is $p = (v_8, v_0, v_4, v_4)$. \deleted[id={I38}]{We allow length\hyp{}one paths, e.g. $p = (v_4)$, and even empty paths.}

\deleted{For example, $p = (v_0, v_1)$ is a subpath of $q = (v_8, v_0, v_1, v_2)$.}

A \emph{test case} (see \ref{sec:graphs:testcases}) is a non\hyp{}empty path starting from the entry vertex. For example, $t_1 = (v_0)$ and $t_2 = (v_0, v_6)$ are test cases but $t_3 = (v_6)$ and $t_4 = ()$ are not. An SUT automatically executes a test case by triggering the behaviors in the order of the test case's vertex sequence. Then, every trigger, i.e., vertex, is equivalent to a \emph{test step} \deleted{(see Appendices)}, which takes some time to execute. A \emph{test suite} \deleted{(see Appendices)} is a collection of test cases, e.g., $T = \lbrace t_1, t_2 \rbrace$. In MBT practice, developers or testers generate and execute test suites to ensure the software quality.

For example, suppose $r$ is the splice of $p = (v_0)$ and $q = (v_5)$.
Then $r$ could either be $(v_0, v_3, v_5)$ or $(v_0, v_4, v_5)$.
The splice can be shorter than the sum of the spliced paths' lengths, e.g., the splice of $(v_0, v_7, v_8)$ and $(v_7, v_8, v_0)$ is $(v_0, v_7, v_8, v_0)$. 
Finally, the splice may not exist, e.g., for $(v_0, v_6)$ and $(v_7, v_8)$. With the help of a shortest path algorithm, computing a splice is straightforward (see \ref{sec:computations:splice}).

\begin{elstable}
    \caption{T\added[id={I10}]{he t}est requirements\added[id={I13}]{ ($\ltrl{TR}$)} \replaced{of}{for} \replaced{the VC, EC, EPC, and EPC}{vertex, edge, edge-pair, and prime path criteria}\added{, derived for the graph model in \cref{fig:examplegraph}}.}
    \label{tab:testreq}
    %\rowcolors{1}{White}{Gray!15}
    \begin{tabular}{rrrrr}
        \toprule
        \# & \multicolumn{1}{c}{$\crit{vc}$} & \multicolumn{1}{c}{$\crit{ec}$} & \multicolumn{1}{c}{$\crit{epc}$} & \multicolumn{1}{c}{$\crit{ppc}$} \\
        \midrule
        $\ltrl{tr}_1$ & $(v_0)$ & $(v_8, v_0)$ & $(v_7, v_8, v_0)$ & $(v_0, v_7, v_8, v_0)$ \\
        $\ltrl{tr}_2$ & $(v_1)$ & $(v_0, v_1)$ & $(v_8, v_0, v_1)$ & $(v_1, v_2, v_1)$ \\
        $\ltrl{tr}_3$ & $(v_2)$ & $(v_2, v_1)$ & $(v_1, v_2, v_1)$ & $(v_2, v_1, v_2)$ \\
        $\ltrl{tr}_4$ & $(v_3)$ & $(v_1, v_2)$ & $(v_0, v_1, v_2)$ & $(v_7, v_8, v_0, v_1, v_2)$ \\
        $\ltrl{tr}_5$ & $(v_4)$ & $(v_0, v_3)$ & $(v_2, v_1, v_2)$ & $(v_4, v_4)$ \\
        $\ltrl{tr}_6$ & $(v_5)$ & $(v_0, v_4)$ & $(v_8, v_0, v_3)$ & $(v_7, v_8, v_0, v_3, v_5)$ \\
        $\ltrl{tr}_7$ & $(v_6)$ & $(v_4, v_4)$ & $(v_8, v_0, v_4)$ & $(v_7, v_8, v_0, v_4, v_5)$ \\
        $\ltrl{tr}_8$ & $(v_7)$ & $(v_3, v_5)$ & \cellcolor{GraphFrameBG}$(v_0, v_4, v_4)$ & $(v_7, v_8, v_0, v_6)$ \\
        $\ltrl{tr}_9$ & $(v_8)$ & $(v_4, v_5)$ & $(v_0, v_3, v_5)$ & $(v_7, v_8, v_0, v_7)$ \\
        $\ltrl{tr}_{10}$ & & $(v_0, v_6)$ & $(v_0, v_4, v_5)$ & $(v_8, v_0, v_7, v_8)$ \\
        $\ltrl{tr}_{11}$ & & $(v_0, v_7)$ & $(v_4, v_4, v_5)$ & \\ 
        $\ltrl{tr}_{12}$ & & $(v_7, v_8)$ & $(v_8, v_0, v_6)$ & \\
        $\ltrl{tr}_{13}$ & & & $(v_8, v_0, v_7)$ & \\
        $\ltrl{tr}_{14}$ & & & $(v_0, v_7, v_8)$ & \\
        \bottomrule
        \multicolumn{5}{r}{\footnotesize \textbf{Highlighted:} An edge pair that is not a subpath of any $\ltrl{tr} \in \crit{ppc}$.}
    \end{tabular}
\end{elstable}

In \cref{tab:testreq}, we present the test requirements for \replaced{the vertex (VC), edge (EC), edge-pair (EPC), and prime path (PPC)}{these} criteria under the example graph model in \cref{fig:examplegraph}. \replaced{While generating the requirements for VC, EC, and EPC are straightforward, a}{A}lgorithms that generate a minimal set of requirements for \added{the }\replaced{PPC}{prime path criterion} are well\hyp{}known (see \ref{sec:computations:ppc}). According to \cref{tab:testreq}, the example graph in \cref{fig:examplegraph} yields ten test requirements for the \replaced{PPC}{prime path criterion}.\added{ Note that the EPC is stronger than the VC and the EC, but the PPC is not stronger than EPC; the edge-pair $(v_0, v_4, v_4)$ is not a subpath of any PPC test requirement.}

\deleted{Finally, to measure coverage, we calculate the percentage of test requirements that a test suite covers.
For example, suppose $t_1 = (v_0, v_7, v_8, v_0, v_6)$, $t_2 = (v_0, v_4, v_4, v_5)$, and $t_3 = (v_0, v_3, v_5)$.
Then, according to Table 1, the test suite $\ltrl{TR} = \lbrace t_1, t_2, t_3 \rbrace$ gets $78\%$ vertex, $75\%$ edge, $43\%$ edge\hyp{}pair, and $30\%$ prime path coverage.}

\deleted{A path is a \emph{subpath} (see Appendices) of another only if both paths belong to the same graph and the latter contains the former. For example, $p = (v_0, v_1)$ is a subpath of $q = (v_8, v_0, v_1, v_2)$.}

\deleted{Notice that every path is a subpath of itself. To eliminate this case, we say that a path is a \emph{strict subpath} (see Appendices) of another path only if the former is a subpath of the latter and the latter is longer than the former.}

\subsection{The DJ method}

\deleted[id={I27}]{We define the \emph{minimal test generation problem} in MBT with graph models as the problem of generating the minimal test suite that achieves $100\%$ coverage for a given criterion.}

\added{There are two key strategies behind the test case minimization of the DJ method, splicing of the test requirements and flow minimization, which covers these splices with the as few test cases as possible.} A \emph{splice} (see \ref{sec:graphs:seqops}) of two paths is a shortest path such that,
\begin{enumerate}
    \item Starts with the first vertex of the former,
    \item Ends with the last vertex of the latter, and
    \item Both paths are subpaths of the splice.
\end{enumerate}

\begin{elsfigure}
    \begin{graphframe}
        \includegraphics{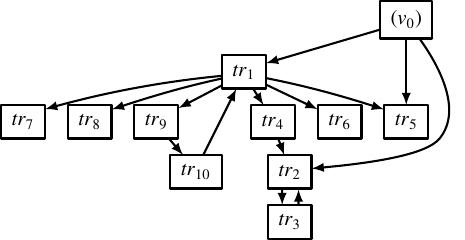}
    \end{graphframe}
    \caption{\replaced[id={I13}]{The splice graph of the PPC test requirements from \cref{tab:testreq}}{Example splice graph for prime path requirements}.}
    \label{fig:expprg}
\end{elsfigure}

\cref{fig:expprg} illustrates the splice graph for the prime path criterion ($\ltrl{TR}_{\text{ppc}}$) in \cref{tab:testreq} and the example graph model in \cref{fig:examplegraph}.
To motivate the strategy behind \replaced{splicing}{this algorithm}, we take a path out of this splice graph, starting from the entry vertex.
Suppose $((v_0), \ltrl{tr}_1, \ltrl{tr}_4, \ltrl{tr}_2)$ is such a path. By chain\hyp{}splicing the sequence of this path, we get
\begin{equation}
    \splice{\splice{\splice{(v_0)}{\ltrl{tr}_1}}{\ltrl{tr}_4}}{\ltrl{tr}_2} = (
    \lefteqn{\underbrace{\phantom{v_0, v_7, v_8, v_0}}_{\ltrl{tr}_1}}v_0, 
    \lefteqn{\overbrace{\phantom{v_7, v_8, v_0, v_1, v_2}}^{\ltrl{tr}_4}}v_7, v_8, v_0, 
    \underbrace{v_1, v_2, v_1}_{\ltrl{tr}_2}
    )\label{eq:exchainsplice}
\end{equation}
which is the shortest test case that covers the selected three of the ten prime path test requirements\deleted[id={I12}]{, achieveing $30\%$ coverage on its own}.

The DJ method comprises the following steps.
\begin{enumerate}
    \item Construction of a splice graph of test requirements.
    \item Construction of an acyclic hypergraph from the splice graph.
    \item Constrained flow minimization of the hypergraph, which minimizes the number of test cases.
    \item Unwinding hyperpaths to test cases.
\end{enumerate}

\added[id={I12}]{We already discussed the first step, which is splice graph construction.\added{ Dwarakanath and Jankiti provide the algorithm for splice graph construction~\cite{DwarakanathJankiti2014:ICTSS}. We also describe this algorithm in \ref{sec:computations:splicegraph}}.} In the following sections, we explain \replaced{the remaining}{these} steps\added{ of DJ}. In the final section, we discuss DJ's suboptimality.

\subsubsection{Hypergraph construction}
\label{sec:back:dj:hypergraph}

\begin{elsfigure}
    \begin{graphframe}
        \includegraphics{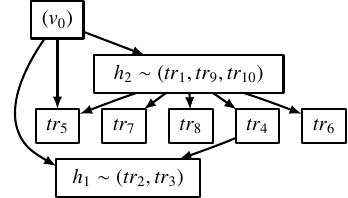}
    \end{graphframe}
    \caption{\replaced[id={I13}]{The acyclic hypergraph generated by iteratively replacing the cycles of the splice graph in \cref{fig:expprg} with hypervertices}{Example hypergraph for prime path requirements}.}
    \label{fig:exhypergraph}
\end{elsfigure}

\replaced[id={I12}]{Since minimum flow optimization is inapplicable over cyclic graphs, the DJ method repeatedly finds cycles in the splice graph and combines them into \emph{hypervertices}, obtaining an acyclic \emph{hypergraph} as in \cref{fig:exhypergraph}.}{Splice graphs facilitate the generation of short test cases, but they do not determine the minimum number of test cases required for $100\%$ coverage. The fact that splice graphs often contain \emph{cycles}, i.e., paths that start from and end at the same vertex (see Appendices), complicates the answer to this question. As a result, the DJ method repeatedly finds a cycle in the splice graph and replaces it with a hyperpath, i.e., a vertex that represents the cycle (see Appendices), until no cycle remains. The resulting acyclic graph (see Appendices) is a \emph{hypergraph} (see Appendices).}

\replaced[id={I16}]{We denote our example hypervertices as $h_1 \sim (\ltrl{tr}_2, \ltrl{tr}_3)$ and $h_2 \sim (\ltrl{tr}_1, \ltrl{tr}_9, \ltrl{tr}_{10})$. Note that these hypervertices are cycles according to \cref{fig:expprg}.}{When refering to a hyperpath, we usually show its prefix, like $h_1 \sim (\ltrl{tr}_2, \ltrl{tr}_3)$ and $h_2 \sim (\ltrl{tr}_1, \ltrl{tr}_9, \ltrl{tr}_{10})$, omitting the target vertex which is equal to the first vertex.}

\subsubsection{Flow minimization}
\label{sec:back:dj:flowmin}

The acyclicity of the hypergraph enables DJ to use a constrained minimum flow variant of the Ford\hyp{}Fulkerson algorithm to minimize the number of test cases~\citep{FordFulkerson1956:CJM}.
The DJ method constrains the minimum flow algorithm to visit \replaced[id={I44}]{every hypervertex}{every vertex of the hypergraph} at least once, so the algorithm does not remove any test requirements. We now describe the essentials of the algorithm, as its intrinsic details are an active subject of research~\citep{Chen+:2025:JACM}.

The minimum flow algorithm performs minimization from a source vertex to a sink. These vertices correspond to the beginning and the end of a test case, respectively. In \replaced[id={I10}]{an}{the} hypergraph, the entry path\added{, a path that comprises only the entry vertex,} represents the beginning, but the end remains implicit; the test case could end \replaced{with}{after} any vertex. Thus, the DJ method creates a new empty \replaced[id={I44}]{hypervertex}{hyperpath} \added[id={I12}]{that represent test termination} and connects every other vertex to it. Then, it assigns a positive integer, called flow, representing the number of test cases that cover it, to every edge. Finally, it minimizes the total flow from the entry path to the empty hyperpath, such that every vertex except the sink in the hypergraph has a positive outgoing flow.

\begin{elsfigure}
    \begin{graphframe}
        \includegraphics{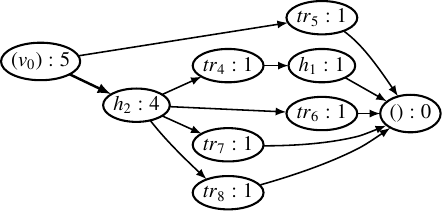}
    \end{graphframe}
    \caption{\replaced[id={I13}]{A minimal flow network from $(v_0)$ to $()$ that traverses every vertex of the hypergraph in \cref{fig:exhypergraph} at least once}{A minimal constrained flow network}.}
    \label{fig:exflow}
\end{elsfigure}

In \cref{fig:exflow}, we present a minimized flow network of the hypergraph in \cref{fig:exhypergraph}. This network's vertices are the same as the hypergraph, with the addition of a sink vertex at the end. Every vertex has a number, corresponding to its total outgoing flow. The sink vertex naturally has zero outgoing flow; it does not have any outgoing edges. All the other vertices have positive total flow. The source has the highest total outgoing flow, five, which is the minimum number of test cases required to cover all test requirements.

Some edges of the hypergraph do not appear in the flow network because they carry zero flow, e.g., $((v_0), h_1)$. As a result, generating five distinct paths from the source to the sink becomes straightforward. A sequence of these paths is
\begin{equation}
    \begin{alignedat}{1}
        \pi_1 = (&((v_0), \ltrl{tr}_5), \\
               &((v_0), h_2, \ltrl{tr}_4, h_1), \\
               &((v_0), h_2, \ltrl{tr}_6), \\
               &((v_0), h_2, \ltrl{tr}_7), \\
               &((v_0), h_2, \ltrl{tr}_8))
    \end{alignedat}\label{eq:exhyperpathseq}
\end{equation}
where we omit the sink since it represents an empty path. Every vertex of the network occurs as many times as its total outgoing flow number in these paths. The DJ method removes the extraneous occurrences of these vertices\replaced[id={I43}]{ because the splice of their predecessor and successors will contain all the necessary test steps and traversing every step of the same test requirement would be redundant. Naturally, the DJ method never removes the extraneous occurences of the source vertex since all test cases must start from the entry vertex. In the end, DJ}{, except the source, from these paths, and} obtains five sequences as
\begin{equation}
    \begin{alignedat}{1}
        \pi_2 = (&((v_0), \ltrl{tr}_5), \\
                &((v_0), h_2, \ltrl{tr}_4, h_1), \\
                &((v_0), \ltrl{tr}_6), \\
                &((v_0), \ltrl{tr}_7), \\
                &((v_0), \ltrl{tr}_8))
    \end{alignedat}\label{eq:exhyperseqreduced}
\end{equation}
where the multiple occurences of the source vertex (the entry path) is unavoidable because this is the minimum number of test cases required for \replaced[id={I27}]{the prime path criterion}{$100\%$ coverage}.

\subsubsection{Unwinding}

\begin{elsalgorithm}[t]
    \caption{Unwinding hyperpaths}\label{alg:unwinding:informal}
    \KwIn{A splice graph, a hypergraph, and a sequence of \replaced[id={I44}]{hypervertex}{hyperpath} sequences}
    \KwOut{A sequence of paths of the splice graph}
    \For{every \replaced[id={I44}]{hypervertex}{hyperpath} sequence}{
        Add the sequence to the sequence of paths.\;
        \If{the sequence contains an \replaced[id={I44}]{hypervertex}{hyperpath}}{
            Replace the \replaced[id={I44}]{hypervertex}{hyperpath} with its cycle. \;
            \Repeat{the sequence is the shortest path}{
                Rotate and splice the cycle between the predecessor and the successor of the \replaced[id={I44}]{hypervertex}{hyperpath}.\;
            }
        }
    }
\end{elsalgorithm}

\emph{Unwinding} refers to the production of test cases from the \replaced[id={I44}]{hypervertex}{hyperpath} sequences coming from the flow network. If these sequences contained no \replaced[id={I44}]{hypervertices}{hyperpaths} and only test requirements, obtaining test cases is only a matter of chain\hyp{}splicing, as in \cref{eq:exchainsplice}. Then, unwinding boils down to reconstructing the cycles represented by \replaced[id={I44}]{hypervertices}{hyperpaths}.

We present the steps of the DJ method to unwind \replaced[id={I44}]{hypervertex}{hyperpath} sequences in \cref{alg:unwinding:informal}. This algorithm traverses every element of each sequence and replaces every \replaced[id={I44}]{hypervertex}{hyperpath} with its cycle.\added[id={I12}]{ Note that the order of hypervertex sequences is arbitrary, and trying every ordering would not minimize the test steps as we discuss in \cref{sec:back:dj:suboptimality}.} Then, it rotates the cycle until the chain\hyp{}splice of its predecessor, itself, and its successor is the shortest.\added[id={I61}]{ One rotation of a cycle moves its last vertex to the front, as defined in \ref{sec:graphs:seqops}. Determining the shortest path is straightforward; DJ rotates the cycle as many times as the cycle's element count, and takes the rotation that yields the shortest splice.}

For the example sequences in \cref{eq:exhyperseqreduced}, \cref{alg:unwinding:informal} replaces $h_2$ and $h_1$, as these are the only vertices foreign to the splice graph in \cref{fig:expprg}.
Suppose the algorithm reconstructs $h_2$'s cycle as $(\ltrl{tr}_9, \ltrl{tr}_{10}, \ltrl{tr}_1)$.
Considering the predecessor $(v_0)$ and the successor $\ltrl{tr}_4$, the algorithm gets the shortest splice after rotating this cycle, like $(\ltrl{tr}_1, \ltrl{tr}_{9}, \ltrl{tr}_{10})$.
Then, the chain\hyp{}splice becomes $((v_0), \ltrl{tr}_1, \ltrl{tr}_{9}, \ltrl{tr}_{10}, \ltrl{tr}_{1}, \ltrl{tr}_{4})$.
Note that since \cref{fig:expprg} does not have an edge between test requirements $\ltrl{tr}_{10}$ and $\ltrl{tr}_{4}$, an extra $\ltrl{tr}_{1}$ appears between them during splicing. As $h_1 \sim (\ltrl{tr}_2, \ltrl{tr}_3)$ has no successor, the algorithm only considers its predecessor ($\ltrl{tr}_4$).

After removing the extraneous occurence of $tr_1$, the resulting sequence is
\begin{equation}
    \begin{alignedat}{2}
        \pi_3 = (&((v_0), \ltrl{tr}_5), \\
                 &((v_0), \ltrl{tr}_1, \ltrl{tr}_9, \ltrl{tr}_{10}, \ltrl{tr}_4, \ltrl{tr}_2, \ltrl{tr}_3), \\
                 &((v_0), \ltrl{tr}_6), \\
                 &((v_0), \ltrl{tr}_7), \\
                 &((v_0), \ltrl{tr}_8))
    \end{alignedat}\label{eq:exunwound}
\end{equation}
which contains every prime path test requirement from \cref{tab:testreq} exactly once.
Finally, through chain\hyp{}splicing, \replaced{DJ}{we} obtain\added{s} the test suite as
\begin{equation}
    \begin{alignedat}{2}
        T = \lbrace &(v_0, v_4, v_4), \\
             &(v_0, v_7, v_8, v_0, v_7, v_8, v_0, v_1, v_2, v_1, v_2), \\
             &(v_0, v_7, v_8, v_0, v_3, v_5), \\
             &(v_0, v_7, v_8, v_0, v_4, v_5), \\
             &(v_0, v_7, v_8, v_0, v_6) \rbrace \\
    \end{alignedat}\label{eq:extestsuite}
\end{equation}
with a total length of $31$ test steps\deleted{, achieving $100\%$ prime path coverage}.

\subsubsection{Suboptimality of DJ}
\label{sec:back:dj:suboptimality}

\begin{elsfigure}
    \begin{subfigure}{0.3\columnwidth}
        \begin{graphframe}
            \includegraphics{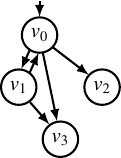}
        \end{graphframe}
        \caption{Graph model $A$.}\label{fig:exsuboptimalA}
    \end{subfigure}
    \begin{subfigure}{0.625\columnwidth}
        \begin{graphframe}
            \includegraphics{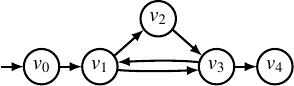}
        \end{graphframe}
        \caption{Graph model $B$.}\label{fig:exsuboptimalB}
    \end{subfigure}
    \caption{Two example graph models\added{.}}
    \label{fig:exsuboptimal}
\end{elsfigure}

In this section, we demonstrate the following causes of suboptimality in the DJ method, all stemming from the \replaced[id={I44}]{hypervertices}{hyperpaths in the graph}.
\begin{enumerate}
    \item Selection of extraneous occurrences of \replaced[id={I44}]{hypervertices}{hyperpaths} and
    \item \replaced[id={I12}]{Prevented}{Prohibited} piecemeal traversal of an \replaced[id={I44}]{hypervertex}{hyperpath}'s cycle, i.e., interrupting the traversal of the cycle in the middle and making a detour before completing the cycle.
\end{enumerate}
We provide two example graph models, $A$ and $B$, in \cref{fig:exsuboptimalA} and \ref{fig:exsuboptimalB}, exposing these suboptimalities, respectively.
In both these examples, we \replaced[id={I12}]{use the VC}{target vertex coverage}.
So, test requirements are length\hyp{}one paths, e.g., $(v_3)$.

\paragraph{For graph $A$}
\replaced[id={I12}]{After processing graph $A$}{After executing algorithms on graph $A$} and minimizing the flow, the DJ method obtains
\begin{equation}
    \begin{alignedat}{2}
        \pi = (&((v_0), h_1, (v_2)),\\
               &((v_0), h_1, (v_3)))
    \end{alignedat}\label{eq:graphAeq1}
\end{equation}
where $h_1 \sim ((v_0), (v_1))$. \replaced[id={I12}]{DJ}{The DJ method} removes one of $h_1$'s occurrences arbitrarily, but the minimality of the test suite depends on which one it selects. For example, if it removes $h_1$ from the second sequence, the resulting test suite is
\begin{equation}
    \begin{alignedat}{2}
        T_1 = \lbrace &(v_0, v_1, v_0, v_2),\\
               &(v_0, v_3) \rbrace
    \end{alignedat}\label{eq:graphAeq2}
\end{equation}
but if it removes $h_1$ from the first sequence, the resulting test suite is
\begin{equation}
    \begin{alignedat}{2}
        T_2 = \lbrace &(v_0, v_2),\\
               &(v_0, v_1, v_3)\rbrace
    \end{alignedat}\label{eq:graphAeq3}
\end{equation}
which has fewer test steps in total than the test suite in \cref{eq:graphAeq2}.

\paragraph{For graph $B$}
\added[id={I12}]{Even if we consider all orderings of hypervertex sequences, DJ produces suboptimal results because it does not consider the cases where the elements of a hypervertex could have been traversed in different parts of a test suite. For example, }
\replaced{after processing}{After executing algorithms on} graph $B$ and minimizing \replaced{its}{the} flow, \replaced{DJ}{the DJ method} obtains
\begin{equation}
    \pi_1 = (((v_0), h_2, (v_4)))\label{eq:graphBeq1}
\end{equation}
where $h_2 \sim (h_1, v_2)$. Unwinding $h_2$, \replaced{DJ}{the DJ method} gets
\begin{equation}
    \pi_2 = (((v_0), h_1, (v_2), h_1, (v_4)))\label{eq:graphBeq2}
\end{equation}
where $h_1 \sim (v_1, v_3)$. Then, \replaced{DJ}{the DJ method} removes one of the extraneous occurrences of $h_1$ to obtain
\begin{equation}
    \pi_3 = (((v_0), h_1, (v_2), (v_4))),\label{eq:graphBeq3}
\end{equation}
performs one more iteration of unwinding to produce
\begin{equation}
    \pi_4 = (((v_0), (v_1), (v_3), (v_1), (v_2), (v_4))),\label{eq:graphBeq4}
\end{equation}
and finally removes the extraneous occurence of $(v_1)$ as
\begin{equation}
    \pi_5 = (((v_0), (v_1), (v_3), (v_2), (v_4))).\label{eq:graphBeq5}
\end{equation}

\deleted[id={I12}]{DJ's unwinding is suboptimal because it prevents piecemeal traversing of vertices in the cycle $(v_1, v_3)$.}

F\deleted{or example, f}rom \cref{eq:graphBeq5}, \replaced{DJ}{the DJ method} generates the test suite
\begin{equation}
    T_1 = \lbrace (v_0, v_1, v_3, v_1, v_2, v_3, v_4) \rbrace\label{eq:graphBeq6}
\end{equation}
where the minimal test suite was
\begin{equation}
    T_2 = \lbrace (v_0, v_1, v_2, v_3, v_4) \rbrace\label{eq:graphBeq7}
\end{equation}
which is impossible for DJ to generate\added{ because the elements of $h_1 \sim (v_1, v_3)$ must be traversed in separate parts of the test case}.

\section{Method}
\label{sec:method}

In the following sections, we explain our novel test generation method, DJPlus, demonstrate how it solves the \replaced{suboptimalities}{problems} presented in \cref{sec:back:dj:suboptimality},\added{ show why DJPlus still fails to generate the absolute minimum number of test steps on an example,} discuss our approach to \replaced{DJPlus'}{the} implementation challenges within the popular GraphWalker environment\added{, and elaborate on its computational complexity}.

\begin{elsfigure}
    \centering
    \includegraphics[trim={1cm 1.25cm 1cm 1.25cm},clip,width=\linewidth]{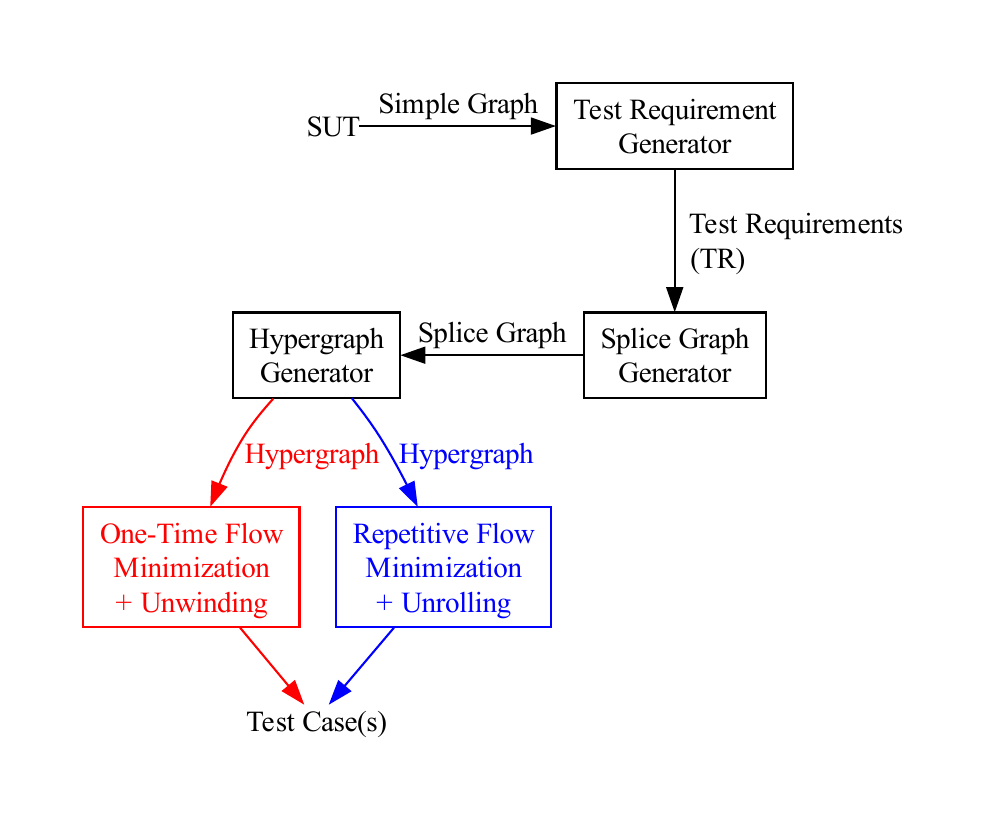}
    \caption{\added[id={I12}]{The differences between the DJ and the DJPlus approaches, common components in black, DJ's components in red, and DJPlus' novel approach is in blue.}}
    \label{fig:djplusdiff}
\end{elsfigure}

\added[id={I12}]{We show an comparative overview of DJ and DJPlus in \cref{fig:djplusdiff}. Both DJ and DJPlus generate an hypergraph in the same way. Then, DJ uses an approach we call \emph{unwinding} to obtain test cases. Unwinding introduces cycles back to the hypergraph, and therefore flow minimization is possible only at the beginning. DJPlus, however, replaces unwinding with a novel approach that we call \emph{unrolling}, which keeps the graph acyclic and allows flow minimization at every iteration.}

\subsection{DJPlus}
\label{sec:method:djplus}

\replaced[id={I12}]{Our general strategy to reduce the number of test steps is to eliminate all hypervertices before generating the test cases, obtaining a flow graph free of hypervertices.}{Since DJ's suboptimalities stem from the hyperpaths in the flow graph, to eliminate these suboptimalities is to obtain a flow graph free of hyperpaths.} We call such a graph a \emph{requirement flow graph} ($\ltrl{RFG}$) because all the vertices are test requirements, except the source and the sink vertices.

\begin{elsalgorithm}[t]
    \caption{Unrolling hyperpaths}\label{alg:unrolling:informal}
    \KwIn{A flow network}
    \KwOut{A requirement flow graph ($\ltrl{RFG})$}
    \For{every \replaced{hypervertex}{hyperpath} $h_i$\deleted{$ \sim (h[0], \ldots, h[\len{h}{\shortminus}1])$}\added[id={I45}]{ \textbf{from} $i=n$ \textbf{to} $1$}}{
        Add \replaced[id={I19}]{$h_i$'s cycle}{$\lbrace h[0], \ldots, h[\len{h}{\shortminus}1] \rbrace$} to $\ltrl{RFG}$.\label{alg:unrolling:informal:unroll} \;
        Add \replaced[id={I19}]{an apostrophe ($'$) decorated vertex for every vertex in $h_i$'s cycle}{$\lbrace h'[0], \ldots, h'[\len{h}{\shortminus}1] \rbrace$ to $\ltrl{RFG}$}.\label{alg:unrolling:informal:unroll2} \;
        Connect every $h[j]$ and $h'[j]$ to the sink. \;
        Reconstruct the edges from and to the \replaced{hypervertex}{hyperpath}'s cycle. \;
        Reconstruct the edges within the \replaced{hypervertex}{hyperpath}'s cycle\added{ without completing the cycle}. \;
        Remove $h_i$ and its edges. \;
        Re-minimize the flow. \;
    }
\end{elsalgorithm}

Our novel strategy to generate an $\ltrl{RFG}$ is to replace every \replaced[id={I44}]{hypervertex}{hyperpath} with its cycle, one at a time, and re-minimize the flow after every replacement. Re-minimization of the flow was normally impossible because replacing a \replaced[id={I44}]{hypervertex}{hyperpath} with its cycle makes the graph cyclic\deleted[id={I12}]{, and flow minimization requires an acyclic graph}. In \cref{alg:unrolling:informal}, we present an unrolling procedure that avoids introducing cycles while reconstructing the elements of \replaced[id={I44}]{hypervertices}{hyperpaths}.

\cref{alg:unrolling:informal} unrolls the cycle of the hyperpath twice in lines~\ref{alg:unrolling:informal:unroll} and \ref{alg:unrolling:informal:unroll2}.
The double unrolling means that now every element of the \replaced[id={I44}]{hypervertex}{hyperpath} appears twice on the flow graph. The algorithm reconnects every edge from and to these unrolled elements but avoids making connections that form a cycle. Thus, it preserves the acyclicity of the flow network. Finally, it re-minimizes the flow, with the following constraints.
\begin{enumerate}
    \item The sink vertex and every second occurrence of the unrolled elements, decorated by an apostrophe \added[id={I12}]{($'$)}, must have zero or more outgoing flow.
    \item The rest must have at least one outgoing flow.
\end{enumerate}

\cref{alg:unrolling:informal} terminates when there are no undecorated \replaced[id={I44}]{hypervertices}{hyperpaths} left to unroll. Note that the algorithm does not unroll decorated \replaced[id={I44}]{hypervertices}{hyperpaths}.

\added[id={I20}]{Notice that we unroll hypervertices in the reverse order of their creation, which ensures every selected hypervertex is not hidden in another hypervertex at the time of unrolling.} We present the \replaced[id={I12}]{other details regarding}{formal steps for} this algorithm in \ref{sec:computations:unrolling}.

\subsection{Motivating examples}
\label{sec:method:examples}

We now demonstrate how DJPlus, unlike DJ, generates \deleted[id={I12}]{minimal }test suites\added[id={I12}]{ with fewer steps} for the example graphs $A$ and $B$ depicted in \cref{fig:exsuboptimal}.

\begin{elsfigure}
    \begin{subfigure}{.4\columnwidth}
        \begin{graphframe}
            \includegraphics{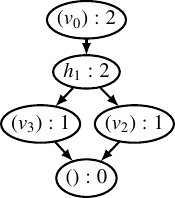}
        \end{graphframe}
        \caption{Graph $A$'s flow network.}\label{fig:hyperA}
    \end{subfigure}
    \begin{subfigure}{.525\columnwidth}
        \begin{graphframe}
            \includegraphics{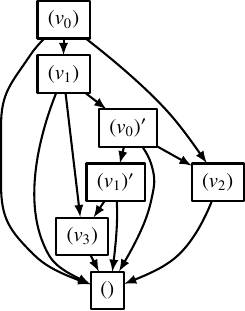}
        \end{graphframe}
        \caption{After unrolling $h_1$.}\label{fig:unrollA}
    \end{subfigure}\\[2pt]
    \begin{subfigure}{.8\columnwidth}
        \begin{graphframe}
            \includegraphics{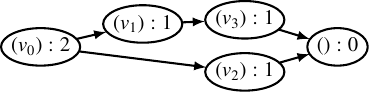}
        \end{graphframe}
        \caption{The $\ltrl{RFG}$ after minimization.}\label{fig:rfgA}        
    \end{subfigure}
    \caption{$\ltrl{RFG}$ generation on graph $A$ from \cref{fig:exsuboptimalA}.}
    \label{fig:exmotivateA}
\end{elsfigure}

\paragraph{Graph A} \cref{fig:hyperA} illustrates an automatically generated and minimized flow network for graph $A$ from \cref{fig:exsuboptimalA}\deleted[id={I12}]{, as described in previous sections}. \cref{alg:unrolling:informal} unrolls $h_1$ twice to produce the intermediate graph in \cref{fig:unrollA}. Then, it performs a re-mini\-mization to generate the $\ltrl{RFG}$ in \cref{fig:rfgA}, which automatically gets rid of the second occurrences of $(v_0)$ and $(v_1)$. From this $\ltrl{RFG}$, DJPlus produces the minimal test suite in \cref{eq:graphAeq3}. In contrast, DJ obtained this test suite only when it arbitrarily selected the correct \replaced[id={I44}]{hypervertex}{hyperpath} to remove, which is not always the case.

\begin{elsfigure}
    \begin{subfigure}{.25\columnwidth}
        \begin{graphframe}
            \includegraphics{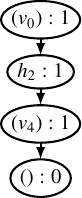}
        \end{graphframe}
        \caption{Graph $B$'s flow.}\label{fig:hyperB}
    \end{subfigure}
    \begin{subfigure}{.45\columnwidth}
        \begin{graphframe}
            \includegraphics{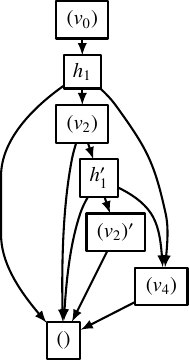}
        \end{graphframe}
        \caption{After unrolling $h_2$.}\label{fig:unrollBh2}
    \end{subfigure}
    \begin{subfigure}{.25\columnwidth}
        \begin{graphframe}
            \includegraphics{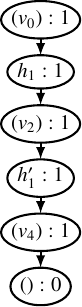}
        \end{graphframe}
        \caption{Minimization {\#}1.}\label{fig:min1B}
    \end{subfigure}\\[2pt]
    \begin{subfigure}{.95\columnwidth}
        \begin{graphframe}
            \includegraphics{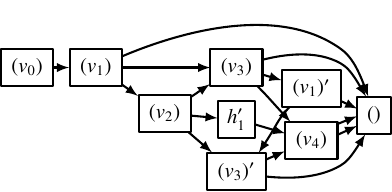}
        \end{graphframe}
        \caption{After unrolling $h_1$.}
        \label{fig:unrollBh1}
    \end{subfigure}\\[2pt]
    \begin{subfigure}{.95\columnwidth}
        \begin{graphframe}
            \includegraphics[scale=.8]{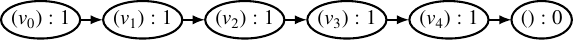}
        \end{graphframe}
        \caption{The $\ltrl{RFG}$ after minimization.}\label{fig:rfgB}
    \end{subfigure}
    \caption{$\ltrl{RFG}$ generation on graph $B$ from \cref{fig:exsuboptimalB}.}
    \label{fig:exmotivateB}
\end{elsfigure}

\paragraph{Graph B} \cref{fig:hyperB} is a minimized flow network for graph $B$ from \cref{fig:exsuboptimalB}. \cref{alg:unrolling:informal} unrolls $h_2$ as in \cref{fig:unrollBh2} and minimizes the flow as in \cref{fig:min1B}. The algorithm managed to remove $h_2$ but introduced another \replaced[id={I44}]{hypervertex}{hyperpath} $h_1$ because the graph model $B$ in \cref{fig:exsuboptimalB} contains nested cycles, i.e., cycles within cycles. So, it unrolls $h_1$ as in \cref{fig:unrollBh1} and then re-minimizes to get the $\ltrl{RFG}$ in \cref{fig:rfgB}. Thus, DJPlus obtains the minimal test suite in \cref{eq:graphBeq7}, which was impossible for DJ.

\subsection{Suboptimality of DJPlus}
\label{sec:method:djplussubop}

\added[id={I21}]{Although DJPlus manages to find shorter test cases than DJ, the number of test steps are still not the absolute minimum. In this section, we step-by-step demonstrate the suboptimality of DJPlus on an example. In this example, we also show how DJPlus handles nested cycles.}

\begin{elsfigure}
    \begin{subfigure}{.45\columnwidth}
        \begin{graphframe}
            \includegraphics{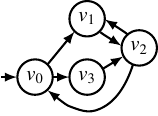}
        \end{graphframe}
        \caption{A graph with nested cycles.}\label{fig:exDJPlusA}
    \end{subfigure}
    \begin{subfigure}{.45\columnwidth}
        \begin{graphframe}
            \includegraphics{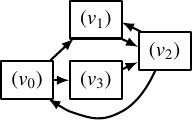}
        \end{graphframe}
        \caption{The splice graph of \cref{fig:exDJPlusA}.}\label{fig:exDJPlusB}
    \end{subfigure}\\[2pt]
    \begin{subfigure}{.45\columnwidth}
        \begin{graphframe}
            \includegraphics{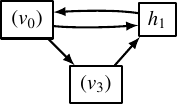}
        \end{graphframe}
        \caption{After combining $h_1 \sim ((v_1), (v_2)))$}\label{fig:exDJPlusC}
    \end{subfigure}
    \begin{subfigure}{.45\columnwidth}
        \begin{graphframe}
            \includegraphics{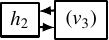}
        \end{graphframe}
        \caption{After combining $h_2 \sim ((v_0), h_1)$}\label{fig:exDJPlusD}
    \end{subfigure}\\[2pt]
    \begin{subfigure}{.9\columnwidth}
        \begin{graphframe}
            \includegraphics[width=\linewidth]{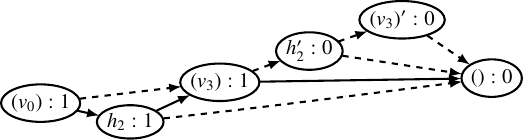}
        \end{graphframe}
        \caption{\added[id={I42}]{Unrolling of $h_3 \sim (h_2, (v_3))$.}}\label{fig:exDJPlusE}
    \end{subfigure}\\[2pt]
    \begin{subfigure}{.9\columnwidth}
        \begin{graphframe}
            \includegraphics{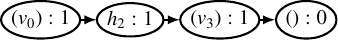}
        \end{graphframe}
        \caption{After removing zero flows.}\label{fig:exDJPlusF}
    \end{subfigure}\\[2pt]
    \caption{\added[id={I20}]{DJPlus's execution on an graph model with nested cycles.}}
    \label{fig:exDJPlus}
\end{elsfigure}

\added[id={I42}]{We present the execution steps of DJPlus on an example graph with nested cycles in \cref{fig:exDJPlus}. DJPlus starts by taking the graph model in \cref{fig:exDJPlusA} as input. Suppose the VC is chosen for this case. Then, the splice graph DJPlus constructed looks very similar to the graph model, as in \cref{fig:exDJPlusB}. DJPlus repeatedly finds the shortest cycle and replaces it with a hypervertex, as in \cref{fig:exDJPlusC,fig:exDJPlusD}. We omit showing the insertion of the final hypervertex, $h_3 \sim (h_2, (v_3))$, as that case is trivial. DJPlus then $(v_0)$ to $()$ that passes through $h_3$,  unrolls $h_3$, and re-minimizes the flow to obtain the graph in \cref{fig:exDJPlusE}. Finally, it removes the zero flows to obtain the graph in \cref{fig:exDJPlusF}. However, in this case, the minimum flow was not unique, the other alternative being $((v_0), (v_3), h'_2, ())$, which also covers all test requirements. DJPlus ignores that, unrolls every hypervertex starting from its chosen flow, and it gets the test case $(v_0, v_1, v_2, v_0, v_3)$, whereas the other choice would have yielded $(v_0, v_3, v_2, v_1)$, which was shorter.}

\added{To make DJPlus optimal, one could conduct a \emph{search} through alternative minimum flows. But, in the worst case, such a search could lead DJPlus to unroll an exponential number of alternatives. In this study, we avoided this case by letting DJPlus select an arbitrary minimum flow and not perform a search. Note that workarounds like changing the selection order of cycles that are combined into hypervertices cannot guarantee one unique minimum flow at every  iteration. So, alternative minimum flows are unavoidable, and are the primary cause of suboptimality in DJPlus. Due to this suboptimality, we opted to empirically evaluate DJPlus' number of test steps and its redundancies.}

\subsection{Implementation challenges}
\label{sec:method:implementation}

We implement DJ and DJPlus for the popular MBT tool, GraphWalker, which features a graphical user interface (GUI) to design graph models, generate a test case\deleted[id={I27}]{ for a given percentage of vertex or edge coverage}, and execute it on the SUT. It features two random test generation algorithms, Random and QRandom. We refer to our implementations on top of GraphWalker as GWPlus.

\deleted{We identify the following challenges of implementing and evaluating DJ and DJPlus for GraphWalker.}

\deleted{1. A GraphWalker model is not a typical graph model. Thus, GWPlus must first convert the model to generate tests.}
\deleted{2. GWPlus must measure edge-pair and prime path coverage percentages.}
\deleted{3. GraphWalker enters an infinite loop when the target coverage criterion is impossible with only one test case.}
\deleted{4. GraphWalker generates tests for varying coverage percentages below $100\%$, whereas DJ and DJPlus always target full coverage, generating redundant test steps.}

\added[id={I12}]{We now discuss practical implementation challenges of our approach within the GraphWalker framework, regarding model conversion, coverage measurements and infinite loops. }We explain how we addressed these issues in the following sections.

\subsubsection{Model conversion}

\replaced[id={I48}]{The popular GraphWalker environment uses graph models similar to the ones described by Dwarakanath and Jankiti, albeit with the following key differences.}{Unlike a typical MBT graph model, a GraphWalker model (see Appendices) has the following properties.}

\begin{enumerate}
    \item \replaced[id={I12}]{A GraphWalker model}{It} comprises many graphs connected by shared vertices\added[id={I12}]{, instead of just one}.
    \item \replaced{Each graph}{Every one of its graphs} is a \emph{multigraph} \deleted{(see Appendices)}, i.e., allows multiple edges between vertices.
    \item \replaced[id={I12}]{The}{Its} entry element \replaced{can be}{is} either a vertex or an edge.
\end{enumerate}

\replaced[id={I12}]{Our practical implementation, GWPlus, takes a GraphWalker model as input and implements both the state-of-the-art DJ and our proposed DJPlus methods. To be able to use the test requirement generation, splice graph construction, hypergraph creation, and the unwinding algorithms of DJ as is, GWPlus employs the strategy to automatically convert GraphWalker models rather than designing variants of these algorithms adapted to GraphWalker.}{From a theoretical standpoint, GraphWalker models are equivalent to MBT graph models. Therefore, our test generation strategy is to convert the GraphWalker model to a graph model.}

GWPlus first creates a unified graph by taking the union of all the vertices and the edges of the GraphWalker model, respectively. The unified graph is still a multigraph, which poses a problem for DJ, DJPlus, and any other deterministic graph-based test generator. Such test generators generate vertex pairs as test requirements for \replaced[id={I10}]{the EC}{edge coverage}, missing some of the extraneous edges between vertices in the process.

GWPlus converts the unified multigraph to a simple graph, where the conversion method depends on the \replaced[id={I12}]{test}{target coverage} criterion. For \replaced[id={I12}]{the VC}{vertex coverage}, removing all extraneous edges is sufficient. For \replaced[id={I12}]{the EC and the EPC}{edge and edge-pair criteria}, GWPlus takes the unified multigraph's \emph{line graph} \deleted{(see Appendices)}, i.e., a graph whose vertices are the edges of the original. For \replaced[id={I10}]{the PPC}{prime path coverage}, GWPlus allows edge removal or line graphs. We name the resulting criteria $\ltrl{prime}_1$ and $\ltrl{prime}_2$ \deleted[id={I12}]{coverage}, respectively.

Line graph conversion affects \replaced[id={I12}]{test criterion}{coverage targets}, e.g., the \replaced[id={I12}]{VC}{vertex coverage} of the line graph is equivalent to the \replaced[id={I12}]{EC}{edge coverage} of the GraphWalker model. Similarly, the \replaced[id={I12}]{EC}{edge coverage} of the line graph is equivalent to the \replaced[id={I12}]{EPC}{edge-pair coverage} of the GraphWalker model. \replaced[id={I10}]{The PPC}{Prime path coverage} of the line graph ($\ltrl{prime}_2$) is stronger than the \replaced[id={I12}]{PPC}{prime path coverage} of the GraphWalker model with extraneous edges removed ($\ltrl{prime}_1$).\added[id={I50}]{ Further discussion on line graphs is available in \ref{sec:linegraphs}.}

The final conversion step addresses the entry element of a GraphWalker model, which is either a vertex or an edge. For \replaced[id={I10}]{the VC}{vertex coverage}, if the entry element is an edge, the entry vertex is the successor of that edge. If the entry element is a vertex, \replaced[id={I48}]{the graph satisfies DJ's requirements with no modifications}{the graph is already simple}. For other criteria, if the entry element is an edge, the entry vertex is the corresponding vertex in the line graph. Otherwise, GWPlus has to add a dummy vertex to represent the entry vertex.

\begin{elsfigure}
    \begin{subfigure}{.45\columnwidth}
        \begin{graphframe}
            \includegraphics[width=.9\linewidth,trim={0 0 0 20mm},clip]{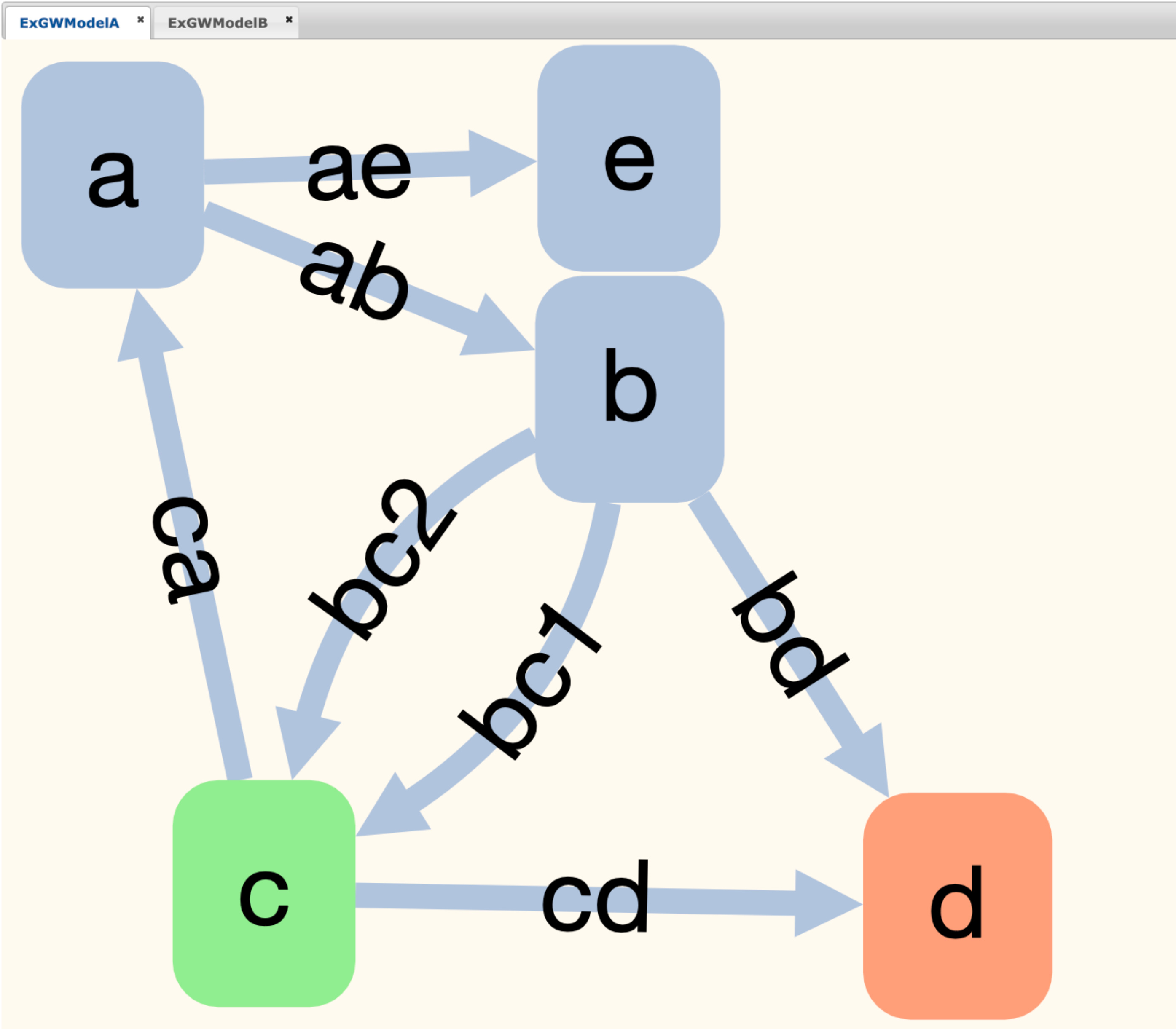}
        \end{graphframe}
        \caption{Multigraph $\ltrl{MG}_A$.}\label{fig:multiA}
    \end{subfigure}
    \begin{subfigure}{.45\columnwidth}
        \begin{graphframe}
            \includegraphics[width=.9\linewidth,trim={0 0 0 20mm},clip]{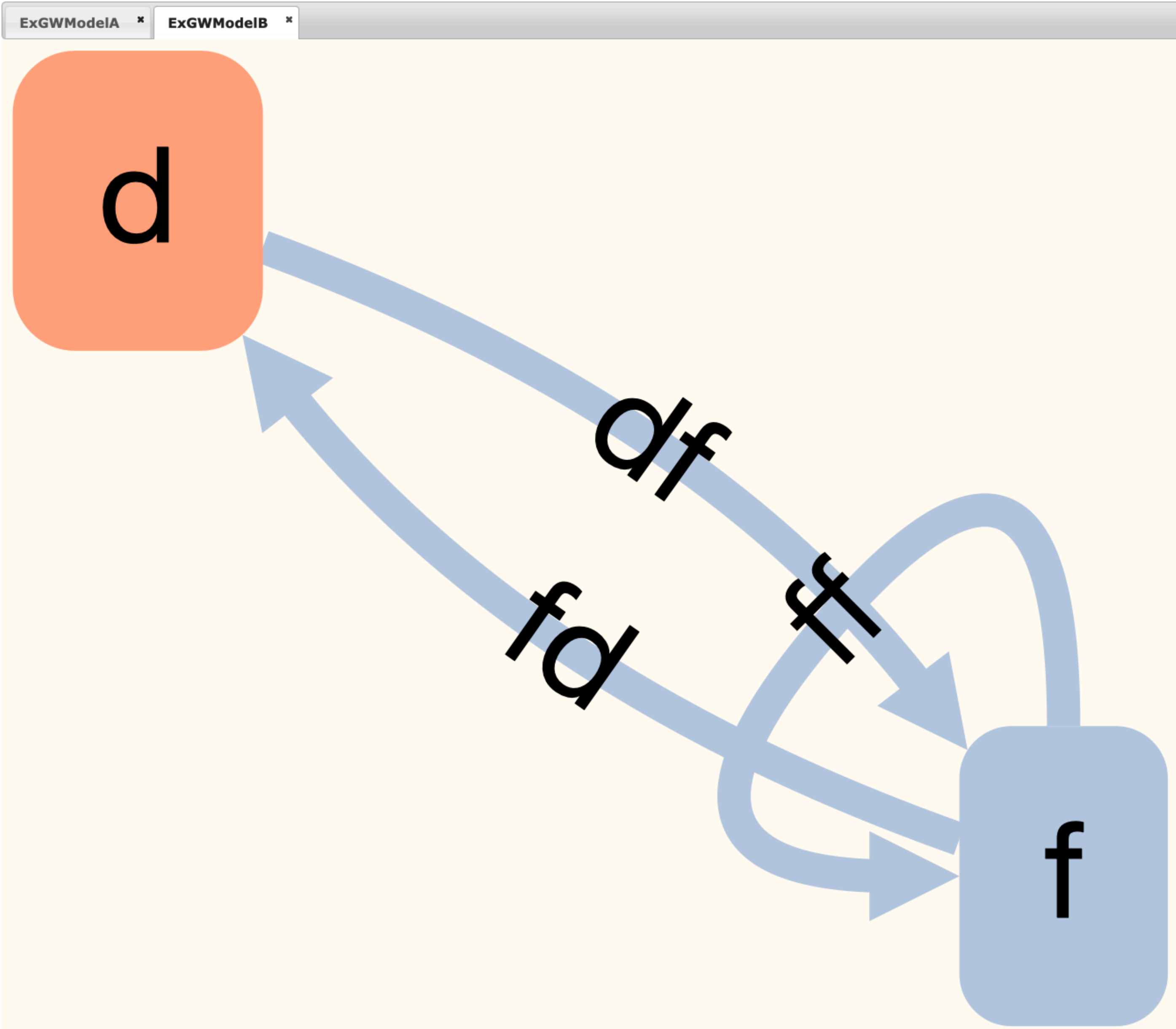}
        \end{graphframe}
        \caption{Multigraph $\ltrl{MG}_B$.}\label{fig:multiB}
    \end{subfigure}\\[2pt]
    \begin{subfigure}{.9\columnwidth}
        \begin{graphframe}
            \includegraphics{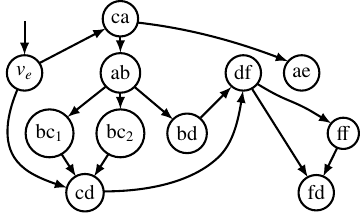}
        \end{graphframe}
        \caption{Unified line graph of the GraphWalker model.}\label{fig:exlinegraph}
    \end{subfigure}
    \caption{\replaced[id={I10}]{An e}{E}xample GraphWalker model with two multigraphs.}\label{fig:exgwmodel}
\end{elsfigure}

\cref{fig:multiA,fig:multiB} depict screenshots of a GraphWalker model's multigraphs we designed in GraphWalker's GUI to demonstrate our conversion procedure. The green colored vertex $c$ in \cref{fig:multiA} is the entry element of this model. The orange colored vertex $d$ is shared by both multigraphs of the model. Notice that there are two edges from $b$ to $c$.

GWPlus generates a unified line graph from the GraphWal\-ker model as in \cref{fig:exlinegraph}, where $v_e$ is the dummy entry vertex. This line graph is \added[id={I12}]{a }simple\added[id={I12}]{ graph} thanks to the uniqueness of its edges.  So, GWPlus moves on to test generation.

\subsubsection{Coverage measurements}

To evaluate coverage\deleted[id={I12}]{performance}, we must first measure it. GraphWalker already measures \replaced[id={I12}]{VC and EC}{vertex and edge coverage percentages}, but does not support the other criteria. GWPlus measures \replaced[id={I12}]{VC, EC, EPC, and PPC}{coverage} by calculating the percentage of \added[id={I12}]{corresponding }test requirements that are subpaths of at least one test case within the generated test suite.

\subsubsection{Infinite loops}

GraphWalker's Random method starts from the entry element and concatenates random successors until the generated test case \replaced[id={I12}]{alone satisfies the target criterion}{achieves a target coverage percentage}. The QRandom method is similar, but splices random unvisited elements instead of concatenating random successors. Sometimes, the target \replaced[id={I27}]{criterion}{percentage} is \replaced[id={I12}]{unsatisfiable}{impossible to achieve} with only one test case\deleted[id={I27}]{, for example, $100\%$ vertex coverage for the GraphWalker model in our example GraphWalker models}. Then, both Random and QRandom enter an infinite loop\deleted[id={I27}]{, causing the tester to terminate test generation manually, albeit at lower coverage}.

To avoid infinite loops, the designer must add extra edges to the model. For this purpose, connecting some vertices back to the entry vertex is typical, representing a restart of the SUT. Note that none of the realistic GraphWalker models we used in our experiments required extra edges to avoid infinite loops.

\noindent\deleted[id={I27}]{\textit{4.3.4. Variable coverage percentages}}

\deleted[id={I27}]{GraphWalker allows targeting a coverage percentage below $100\%$. However, DJ and DJPlus generate test suites that achieve full coverage, regardless of the target percentage. To mitigate the resulting redundancy, GWPlus performs a straightforward post-reduction of the generated test suite. It removes the latest generated test case from the test suite until test case removal becomes impossible without dropping below the target percentage. Then, it removes the final test step of the latest test case until the same condition.}

\subsection{Computational complexity of unrolling in DJPlus}
\label{sec:method:bigoh}

\begin{elstable}
    \centering
    \caption{\added[id={I14}]{The worst-case time complexities of DJPlus algorithms.}}
\label{tab:complexities}
\begin{tabular}{|r|r|}
    \hline
    Splice Graph Construction & $O(\len{\ltrl{TR}}^3)$ \\
    Hypergraph Construction & $O(\len{H}\len{\ltrl{TR}})$\\
    Unrolling & $O(\len{H}\len{\ltrl{TR}}^2)$\\
    \hline
    \multicolumn{2}{r}{\footnotesize $\len{H}$: The number of hypervertices, i.e., cycles in the splice graph.}\\[-2pt]
    \multicolumn{2}{r}{\footnotesize $\len{\ltrl{TR}}$: The number of test requirements.}
\end{tabular}
\end{elstable}

\added[id={I14}]{In this section, we discuss the time complexities of DJ and DJPlus. Dwarakanath and Jankiti provide a rigorous analysis for DJ, and show that the computational complexity is dominated by the splice graph construction~\citep{DwarakanathJankiti2014:ICTSS}. According to their work, assuming $\len{\ltrl{TR}} > \len{\ltrl{E}}$, the worst-case time complexity is $O(\len{\ltrl{TR}}^2 \max(\len{\ltrl{TR}}, \len{\ltrl{E}})) = O(\len{\ltrl{TR}}^3)$, i.e., proportional to the cube of the number of test requirements.}

\added{We derive the worst-case time complexity of hypergraph construction as follows. For every hypervertex, i.e., cycle in the splice graph, the algorithm finds that cycle and replaces it. The complexity of finding a cycle is $\len{V_s} + \len{E_s}$, where the vertices of the splice graph is the set of test requirements ($V_s = \ltrl{TR}$). Assuming that $\len{V_s}$ is linearly proportional to $\len{E_s}$, the overall complexity could be written as $O(\len{H}\len{TR})$, where $\len{H}$ stands for the number of hypervertices, i.e., cycles in the splice graph.}

\added{Finally, DJPlus, on top of DJ, introduces \emph{unrolling}. For every hypervertex, the algorithm unrolls its cycle twice and performs one minimum flow optimization. The minimum flow optimization is tha same as DJ's, whose worst-case complexity is $O(\len{V}\len{E})$. Using our assumptions, we rewrite that complexity as $O(\len{\ltrl{TR}}^2)$. }\added[id={I22}]{Since unlike DJ, DJPlus performs flow minimization not just once but for every hypervertex, the overall complexity becomes $O(\len{H}\len{\ltrl{TR}}^2)$.}

\added{We summarize the worst-case time complexities we analyzed so far for DJPlus in \cref{tab:complexities}. From this table, the time required for unrolling clearly dominates the time needed for hyperpath construction. However, it remains unclear if splice graph construction remains the bottleneck as in DJ, or unrolling is costlier.}

\added{A study published three years after DJ shows that, in the worst-case, a graph comprises $1.44^{\len{E}}$ cycles~\citep{ArmanTsaturian2017:arXiv}. Using this information, the worst-case time complexity of hypergraph construction and unrolling would become exponential. With the absence of a polynomial flow minimization technique directly applicable to cyclical graphs, the only way to keep the complexity of DJ and DJPlus polynomial is to keep the number of cycles low by design of the model. We analyze the experimental SUTs and further discuss our empirical results for test generation times in \cref{sec:eval:results:p2}.}

\section{Evaluation}\label{sec:evaluation}

\replaced[id={I52}]{In the following sections, describe our experimental setup, empirical evaluation of our research questions, and final notes on the test generation times we observed during these experiments.}{We now pose research questions and answer them through experimental results in the following sections.}

\subsection{Experimental setup}
\label{sec:eval:setup}

We now discuss the experimental setup we used to answer the research questions.
In this setup, an experimental run has three phases;
\begin{enumerate}
    \item Test requirement generation,
    \item Test generation, and
    \item Test execution,
\end{enumerate}
where GWPlus automatically generate the test requirements, and we use the command\hyp{}line interface of GraphWalker 4.3.2, the latest GraphWalker version at the time, for test generation. The test execution procedure depends on the SUT, so we discuss every SUT's test execution specifics in \cref{sec:eval:setup:suts}.

Every variable of an experimental run is an experimental setting. In our experimental setup, we have \replaced[id={I27}]{four}{five} experimental settings.
\begin{enumerate}
    \item SUT (TLC, RISC-V, Parabank, or Testinium).
    \item \replaced{Test}{Target} criterion (\replaced{VC, EC, EPC, PPC}{vertex, edge, edge\hyp{}pair, or prime path coverage}).
    \item Test generation method (Random, QRandom, DJ, or DJPlus).
    \item Random seed (One of 30 pre\hyp{}generated random positive integers).
\end{enumerate}

An experimental configuration is a combination of experimental settings. Not every one of the $4 \times 4 \times 4 \times 30 = \num{1920}$ combinations is a valid experimental configuration, e.g., Random cannot target edge\hyp{}pair criterion. We now explain our experimental settings in detail and identify some invalid configurations.

\subsubsection{Systems under test (SUTs)}
\label{sec:eval:setup:suts}

For our experimental setup, we found four realistic SUTs with GraphWalker models.
These models have varying sizes ranging up to $129$ vertices.
Two of them are hardware systems, and the other two are web systems.

\paragraph{Hardware systems} \emph{TLC}\footnote{See \url{https://github.com/kilincceker/MBIT4HW}.} is a Verilog traffic light controller for a four\hyp{}way road instersection, featuring a GraphWalker model with ten vertices and $18$ edges \citep{Kilincceker+2022:SSM}.
We executed test cases on TLC through a simulation environment provided by the Vivado Design Suite.
In accordance with the requirements of the Vivado Design Suite, we carried out the simulations on a PC (Intel Core i7\hyp{}7500U, 2 Cores, 8 GB Memory) with a Windows 10 operating system.

\emph{RISC-V}\footnote{See \url{https://github.com/openhwgroup/cv32e40p}.} is a finite\hyp{}state machine Verilog controller of a $32$\hyp{}bit central processing unit developed by the OpenHW Group, with a small GraphWalker model of $16$ vertices and $36$ edges \citep{Gautschi+2017:TVLSI}.
Like in TLC, we used the Vivado Design Suite to execute RISC-V test cases on the same PC.

\paragraph{Web systems}\emph{Parabank}\footnote{See \url{https://github.com/parasoft/parabank}.} is an open\hyp{}source online banking application from Parasoft.
A recent study used it to evaluate a scriptless method that generates tests on the fly \citep{Hufkens+2024:AST}.
It features a sizable GraphWalker model with $75$ vertices and $144$ edges.
For executing Parabank test cases in headless mode of Chromium version 124, we used the Selenium framework with a GraphWalker 4.3.3 snapshot customized to support the web system on a 10\hyp{}core 16 GB Apple M1 Pro with macOS Sonoma 14.5.

\emph{Testinium}\footnote{See \url{https://github.com/vgarousi/MBTofTestinium}.} is a cloud web application for test automation and management developed by the Testinium company.
One study used it to compare MBT tools like TestOptimal and GraphWalker \citep{Garousi+2021:JSS}.
It boasts the largest GraphWalker model in our experimental setup, with $129$ vertices and $259$ edges.
We noticed substantial changes in the Testinium interface and its GraphWalker model became outdated.
So, we were unable to execute Testinium test cases to obtain results for RQ\replaced{4}{5}.

\subsubsection{Target criteria}

Our experimental configurations allow \replaced{the VC, EC, EPC, and PPC}{vertex, edge, edge\hyp{}pair, and prime path criteria}. Random and QRandom support only \replaced{VC and EC}{vertex and edge coverages}, while DJ and DJPlus support all four criteria.

%\subsubsection{Target percentages}

\noindent\deleted[id={I27}]{\textit{5.1.3. Target percentages}}

\deleted{Our experimental target coverage percentages range from $50\%$ to $100\%$ with ten percent increments.}

\noindent\deleted[id={I12}]{\textit{5.1.4. Test generation methods}}

\deleted{Our experimental configurations comprise the popular Random and QRandom methods, the previous state\hyp{}of\hyp{}the\hyp{}art DJ method, and our novel DJPlus method.}

\subsubsection{Random seeds}

Random and QRandom are not deterministic due to their dependence on pseudo\hyp{}random number generators.
For the sake of reproducibility, we fixed $30$ random seed values.
Executing Random or QRandom with these random seeds allows us to average the results to mitigate the random noise while generating the same tests whenever we rerun the experiments.
Seed values are useless for DJ and DJPlus because these methods are entirely deterministic.

\subsection{Empirical results}
\label{sec:eval:results}

In the following sections, we discuss the results from running the three phases of our experimental setup with different configurations, while addressing the research questions in the process.

\subsubsection{Phase 1: Test requirement generation}
\label{sec:eval:results:p1}

\begin{elstable}
    \caption{\replaced[id={I10}]{The n}{N}umber of test requirements for SUT-criterion pairs, generated by GWPlus.}
    \label{tab:trsizes}
    \begin{tabular}{r|r|r|r|r|r}
        \hline
        SUT & VC & EC & EPC & $\ltrl{prime}_1$ & $\ltrl{prime}_2$ \\
        \hline
        TLC & $10$ & $18$ & $27$ & $60$ & $134$ \\
        RISC-V & $16$ & $36$ & $115$ & $690$ & $>1$M \\
        Parabank & $75$ & $144$ & $501$ & $\num{7457}$ & $>1$M \\
        Testinium & $129$ & $259$ & $\num{1033}$ & $>1$M & $>1$M \\
        \hline
        \multicolumn{6}{r}{\footnotesize VC: vertex criterion, EC: edge criterion, EPC: edge-pair criterion}
    \end{tabular}
    % \pgfplotstabletypeset[
    %     column type={r},
    %     sort,sort key=VC,sort cmp=float <,
    %     columns/SUT/.style={string type},
    %     columns/PPC1/.style={column name={$\ltrl{prime}_1$}},
    %     columns/PPC2/.style={column name={$\ltrl{prime}_2$}},
    %     every column/.style={
    %         postproc cell content/.append code={
    %             \ifnum\pgfplotstablecol>0
    %                 \ifnum##1>999999\pgfkeyssetvalue{/pgfplots/table/@cell content}{$>\!1$M}\fi
    %             \fi
    %         }
    %     },
    %     every last row/.style={after row={\bottomrule\multicolumn{6}{r}{\footnotesize VC: vertex criterion, EC: edge criterion, EPC: edge-pair criterion}}}
    % ]{tables-trsizes.csv}
\end{elstable}

We present the number of GWPlus\hyp{}generated test requirements for our experimental SUTs and target criteria in \cref{tab:trsizes}\deleted[id={I12}]{, where we abbreviate vertex, edge, and edge\hyp{}pair criteria as VC, EC, and EPC, respectively}. Note that the number of test requirements for VC and EC is the same as the number of vertices and edges of the SUT model.

\added[id={I26}]{}From \cref{tab:trsizes}, as the model grows in size, we observe that the number of test requirements for \replaced[id={I10}]{the PPC}{prime path coverage} exceeds a million, which is a significant scalability problem for the \replaced[id={I12}]{PPC}{prime path criterion}, whereas \added[id={I10}]{the }EPC remained within comparatively reasonable boundaries. Thus,
%\begin{qblock}
 \replaced[id={I10}]{the PPC}{prime path criterion} does not scale well, while the other criteria are relatively more scalable.
Therefore, we discard the \replaced{PPC}{prime path criterion} setting for the rest of our experiments.

\subsubsection{Phase 2: Test generation}
\label{sec:eval:results:p2}

\begin{elsfigure}
    \includegraphics{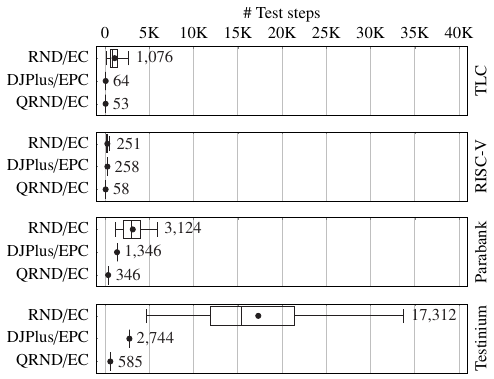}
    \caption{\replaced{The n}{N}umber of test steps by Random (RND) and QRandom (QRND) for \replaced{EC and DJPlus for EPC}{$100\%$ edge coverage (EC), and DJPlus for $100\%$ edge-pair coverage (EPC)}\added[id={I54}]{, where the numbers next to the boxplots show the average values}.}\label{fig:rq2}
\end{elsfigure}

\cref{fig:rq2} shows, with boxplots, for all the four SUT models from the smallest to the largest, \added{the }distributions of the number of steps of three experimental configurations. These configurations are Random for \replaced{the EC (RND/EC)}{$100\%$ edge coverage (RND/EC/100)}, \replaced{DJPlus for the EPC}{DJPlus for $100\%$ edge\hyp{}pair coverage (DJPlus/EPC/100)}, and \replaced{QRandom for the EC (QRND/EC)}{QRandom for $100\%$ edge coverage (QRND/EC/100)}. The boxplots represent the distribution of the number of test steps for all $30$ seed values, with the numbers adjacent to the boxplots indicating the average, denoted by a bullet within the box.

According to \cref{fig:rq2}, Random's variance in the number of test steps grows as the model size grows. DJPlus has zero variance because it is a deterministic method. So, the average of DJPlus is the number of test steps it always generates. Surprisingly, QRandom's variance is nonzero but almost negligible compared to Random.

\paragraph{RQ\replaced{1}{2}}
We consider \replaced[id={I10}]{the PPC}{prime path criterion} to be \emph{infeasible} because it is not scalable as described in \cref{sec:eval:results:p1}. A recent study stated that \replaced[id={I10}]{the EC}{full edge coverage} with GraphWalker for Testinium, the largest SUT model in our experiments, is feasible \citep{Garousi+2021:JSS}. \added[id={I4}]{Considering the feasibility of the EPC, we atypically compare the test steps DJPlus generates for the EPC against the test steps generated by the state-of-the-art tools for EC, and argue that any number of test steps that is below the number of test steps used for the EC should be feasible. }Since the DJPlus test suite for the \replaced{EPC}{edge\hyp{}pair criterion} is \replaced{betweeen}{comparable to} Random and QRandom's test suites for the \replaced{EC}{edge criterion},
\begin{qblock}
\replaced{The EPC}{edge\hyp{}pair criterion} is feasible with DJPlus for realistic SUT models.
\end{qblock}
\paragraph{RQ\replaced{2}{3}}

\begin{elsfigure}
    \includegraphics{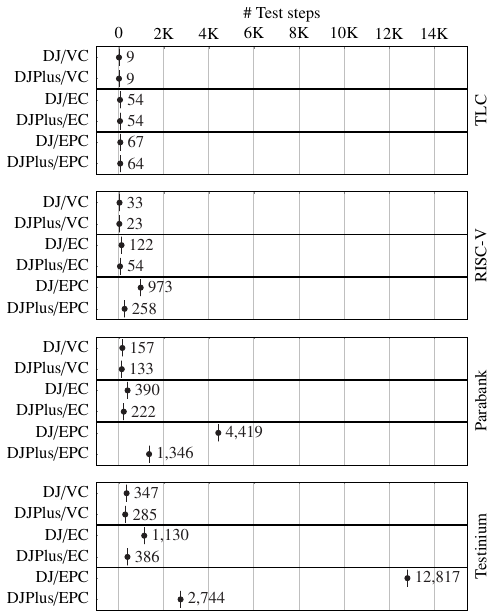}
    \caption{\replaced{The n}{N}umber of test steps by DJ and DJPlus for all \replaced{test criteria}{SUTs}.}\label{fig:rq3}
\end{elsfigure}

We present the number of test steps generated by DJ and DJPlus for all SUTs and \replaced[id={I12}]{test}{target} criteria in \cref{fig:rq3}, where both methods are entirely deterministic and exhibit zero variance. The gap between DJ and DJPlus grows as the model sizes grow. The most significant gap occurs when both methods target \replaced[id={I10}]{the EPC}{edge\hyp{}pair coverage} on Testinium. For all experimental configurations in \cref{fig:rq3}, DJPlus either generated the same number of test steps as or fewer than DJ. Thus,
\begin{qblock}
DJPlus generates test suites with fewer test steps than the state\hyp{}of\hyp{}the\hyp{}art method, DJ.
\end{qblock}
\paragraph{\added{Test Generation Times}}
\begin{elstable}
    \caption{\added[id={I23}]{Average test generation times for all SUTs and test criteria, in seconds.}}
    \label{tab:testgentimes}
    \centering
    \begin{tabular}{r|rrrr}
        \hline
        \multicolumn{5}{c}{\textbf{Vertex Criterion (VC)}} \\
        \hline
          \multirow{2}{*}{\textbf{Method}} & \multicolumn{4}{c}{System Under Test (SUT)} \\
          & TLC & RISC-V & Parabank & Testinium \\
         \hline
         Random & $1.252$ & $1.221$ & $1.450$ & $2.037$ \\
         QRandom & $1.189$ & $1.246$ & $1.315$ & $1.410$ \\
         DJ & $1.168$ & $1.213$ & $1.345$ & $3.078$ \\
         DJPlus & $1.205$ & $1.183$ & $1.375$ & $1.443$ \\
         \hline
         \hline
        \multicolumn{5}{c}{\textbf{Edge Criterion (EC)}} \\
        \hline
          \multirow{2}{*}{\textbf{Method}} & \multicolumn{4}{c}{System Under Test (SUT)} \\
          & TLC & RISC-V & Parabank & Testinium \\
         \hline
         Random & $1.292$ & $1.278$ & $1.633$ & $2.561$ \\
         QRandom & $1.210$ & $1.260$ & $1.378$ & $1.423$ \\
         DJ & $1.186$ & $1.237$ & $1.561$ & $4.401$ \\
         DJPlus & $1.202$ & $1.237$ & $1.625$ & $3.983$ \\
         \hline
         \hline
        \multicolumn{5}{c}{\textbf{Edge-Pair Criterion (EPC)}} \\
        \hline
          \multirow{2}{*}{\textbf{Method}} & \multicolumn{4}{c}{System Under Test (SUT)} \\
          & TLC & RISC-V & Parabank & Testinium \\
         \hline
         DJ & $1.208$ & $2.647$ & $9.866$ & $122.635$ \\
         DJPlus & $1.227$ & $2.605$ & $14.775$ & $186.631$ \\
         \hline
    \end{tabular}
\end{elstable}

\added[id={I23}]{We provide the test generation times of our experiments in \cref{tab:testgentimes}. According to these results, for small models, like TLC and RISC-V, the test generation time of any configuration, regardless of its method or test criterion strength, was below three seconds. However, our largest model, Testinium, when the test criterion is the EPC, required two to three minutes with the available tools. DJPlus particularly appears to have the worst scalability, as it took around 186 seconds for it to generate the Testinium test suite for the EPC. On the other hand, a previous study reported that it took six hours to execute the tests for Testinium~\citep{Garousi+2021:JSS}. So, for our experimental SUTs, we identified the test execution time as the bottleneck, not the test generation time. To comment more on the test generation times, models much larger than our largest, Testinium, are necessary.}

\paragraph{RQ\replaced{3}{4}}
\replaced[id={I56}]{Since Random, QRandom, and DJ all satisfy the same test requirements with more test steps, the utility provided by their extra steps comes into question. We recognize the fact that these extra steps are unnecessary in terms of the test criterion but may still make a difference under a stronger criterion.}{DJPlus minimizes the number of test steps in a test suite while maintaining coverage (see Appendices for further details). Thus, DJPlus must be eliminating some redundancy with respect to a coverage criterion. So, to quantify the redundancy, we must consider both the generated test steps and the targeted test requirements.}

\deleted{In coverage\hyp{}based testing, a coverage criterion typically requires testing an SUT behavior only once. Therefore, we consider any re\hyp{}triggering of an SUT behavior to be redundant.} In light of \replaced[id={I12}]{the above}{this} perspective on redundancy, we define the edge\hyp{}pair redundancy factor of a test suite as
\begin{equation}
    \rf(T) = \frac{\sum_{t \in T, \len{t} > 2} \left(\len{t} - 2\right)}{\cov{epc}(T)\len{\crit{epc}}} - 1 \label{eq:epcrf}
\end{equation}
where the numerator is the total number of edge pairs in the test suite and the denominator is the total number of distinct edge pairs the test suite covers. We calculate the denominator by multiplying the test suite's edge\hyp{}pair coverage ratio\footnote{A coverage ratio is a coverage percentage divided by $100$.\added[id={I55}]{ Multiplying the coverage ratio ($\cov{epc}(T)$) and the total number of edge-pairs ($\len{\crit{epc}}$) gives us exactly the number of distinct edge-pairs covered by $T$.}} ($\cov{epc}(T)$) and the total number of edge pairs in the graph model ($\len{\crit{epc}}$). After subtracting one from the ratio, we obtain a nonnegative redundancy factor ($\rf(T)$). If the redundancy factor is zero, the test suite is optimal, i.e., cannot be further minimized without sacrificing coverage. There is no upper limit to the redundancy factor; a test suite can be many times redundant than the theoretical optimal.

We define the relative redundancy of a test suite ($T$) over another ($U$) as
\begin{equation}
    \rr(T,U) = \deleted{100}\left(\frac{\rf(T)}{\rf(U)} - 1\right).\label{eq:epcrr}
\end{equation}

We express relative redundancy \replaced[id={I12}]{such that it shows how many times}{as a percentage, quantifying how much} more redundant $T$ is than $U$. Unlike the redundancy factor, relative redundancy can be negative, meaning that $T$ is less redundant than $U$.

\begin{elstable}
    \caption{Average redundancy factors ($\rf$) and relative redundancies ($\rr$) over DJPlus.}
    \label{tab:redundancies}
    \begin{tabular}{rrr}
         \hline
         Test Generation Method ($T$) & $\rf(T)$ & $\rr(T,\text{DJPlus})$  \\
         \hline
         Random & $4.32$ & $26.21$x \\
         DJ & $0.49$ & $4.24$x \\
         QRandom & $0.28$ & $2.06$x \\
         DJPlus & $0.17$ & $-$ \\
         \hline
    \end{tabular}
    % \pgfplotstabletypeset[
    %     column type={r},
    %     sort,sort key=RF,sort cmp=float >,
    %     columns/Method/.style={string type,column name={Test Generation Method ($T$)}},
    %     columns/RF/.style={column name={$\rf(T)$},fixed zerofill,precision=2},
    %     columns/RR/.style={column name={$\rr(T,\text{DJPlus})$},fixed zerofill,precision=0,
    %         postproc cell content/.append style={
    %             /pgfplots/table/@cell content/.add={}{\if##1\empty\else\%\fi},
    %         }
    %     }%,
    %     % every last row/.style={after row={\bottomrule\multicolumn{3}{r}{
    %     %     \footnotesize $\rr_{\text{DJPlus}}$: average relative redundancy over DJPlus
    %     % }}}
    % ]{tables-redundancies.csv}
\end{elstable}

We present the average redundancy factors and relative redundancies of Random, DJ, QRandom, and DJPlus in \cref{tab:redundancies}, which we explain further in \ref{sec:redundancies}. According to these results, on average, DJPlus is the least redundant method and
\begin{qblock}
QRandom, DJ, and Random generate between 2 to 26 times more redundant test steps than DJPlus.
\end{qblock}

\begin{elsfigure*}
    \includegraphics{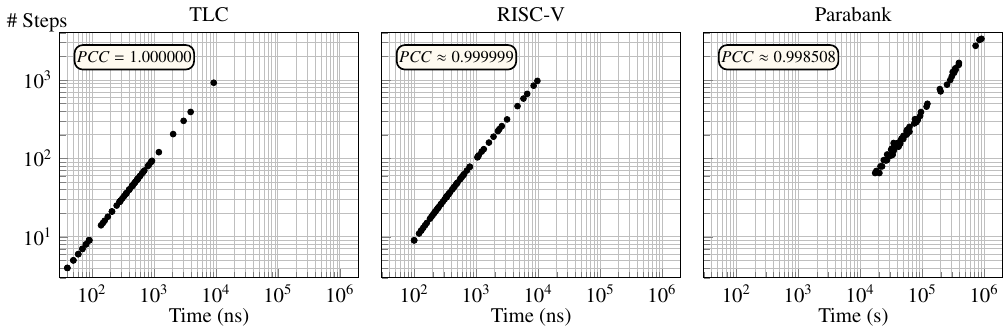}
    \caption{Log-log plots of average test execution time vs. number of test steps for TLC, RISC-V, and Parabank ($\ltrl{PCC}$: Pearson Correlation Coefficient).}\label{fig:rq5}
\end{elsfigure*}

\subsubsection{Phase 3: Test execution}
\label{sec:eval:results:p3}

We present log\hyp{}log plots of the average test execution time versus the number of test steps for TLC, RISC-V, and Parabank in \cref{fig:rq5}, where an average execution time is the average of ten re\hyp{}executions of the same test suite.
To produce these plots, we used the DJ and DJPlus test suites, and the medians of the Random and QRandom test suites for all the percentages of vertex and edge coverage.
Finally, $\ltrl{PCC}$ stands for Pearson Correlation Coefficient\added[id={I3}]{}\added[id={I58}]{, which is an effect size metric}.
A $\ltrl{PCC}$ value close to one indicates a positive linear correlation.

\paragraph{RQ\replaced{4}{5}}
All the $\ltrl{PCC}$ values for TLC, RISC-V, and Parabank indicated a strong linear correlation, meaning that in our experiments, few test steps resulted in low execution times and vice versa.
Thus,
\begin{qblock}
DJPlus decreases the execution times by reducing the number of test steps.
\end{qblock}

\section{Discussion}\label{sec:discussion}

We discuss the issues regarding the impact, redundancy, reproducibility, generalizability, and correlation analysis of our method, DJPlus, and its evaluation in the following sections.

\subsection{Impact}
\label{sec:disc:impact}
The impact of decreasing test execution times depends on the testing practice.
Minimization is not critical if test execution is a one\hyp{}off. The same is not the case with coverage\hyp{}based \nohyphens{testing}, in which developers maintain test cases to ensure software quality and re\hyp{}execute them whenever an SUT is updated.
Thus,
\begin{qblock}
a reduced test suite for strong coverage is highly desirable in coverage\hyp{}based testing, where the adoption of an MBT approach in an industrial\hyp{}scale continuous integration / continuous development process depends on the approach being affordable within a limited testing budget.
\end{qblock}

\subsection{Redundancy}
\label{sec:disc:redundancy}
The topography of a graph model limits the minimum attainable redundancy factor due to \emph{critical components}, which are subgraphs that a test case cannot avoid traversing multiple times to achieve a coverage target. 
\deleted{Edge\hyp{}pair redundancy is preferable to vertex or edge redundancy because, in a graph model, a critical edge pair is less likely to occur than a critical vertex or edge (see Appendices), allowing test generation methods to approach a zero redundancy factor.
Our experimental results confirm that such edge pairs are rare, as DJPlus eliminates a significant amount of redundancy in test generation.}

\added[id={I6}]{}The redundancy of some test steps in a test suite with respect to a coverage criterion does not mean they are redundant for other purposes. 
A good example is stress testing, where a test case repeatedly re\hyp{}executes the same parts of an SUT to wear it down. 
The redundant test steps also have some potential to facilitate fault detection. 
To the best of our knowledge, an evaluation of GraphWalker's coverage\hyp{}redundant test steps with respect to these criteria is an open question in the literature. 
A recent study investigates the possibility of redundant GraphWalker test steps increasing edge\hyp{}pair coverage, finding that the cost of a unit increase in edge\hyp{}pair coverage through redundant GraphWalker steps grows exponentially~\citep{Koroglu+2025:ICSTW:AMOST}.

\subsection{Reproducibility}
\label{sec:disc:reproducibility}
In the first two phases (test requirement and test generation), we control the sources of randomness, so our experiments are fully reproducible.
However, the last phase involves measuring test execution times that depend on the environment and therefore suffer from random noise. 
By re\hyp{}executing the same test ten times, we decreased the effects of random noise. So, even if the exact test execution times we obtained are not reproducible, their trends are replicable.

\subsection{Generalizability}
\label{sec:disc:generalizability}
\added[id={I7}]{}With a limited number of SUTs to evaluate testing methods, the generalizability of our results to an arbitrary SUT is always a question.
To mitigate bias, we selected realistic SUTs with existing models of varying sizes and implementations from two distant domains, hardware and web applications.

\subsection{Correlation}
\label{sec:disc:correlation}
\added[id={I3}]{}In \cref{sec:eval:results:p3}, we show that the number of test steps and the execution time of a test suite are strongly correlated.
We get these results from our controlled experiments, where the number of test steps is the only independent variable, i.e, the only factor that we change to produce the scatter plots in \cref{fig:rq5}.
Thus,
\begin{qblock}
reducing test steps \enquote{causes} a decrease in test execution times.
\end{qblock}

\subsection{Graph coverage}

\added[id={I18}]{The test criteria we use throughout this paper correspond to metrics that measure the coverage of the model and not the implementation of the SUT. Therefore, the increased strength that comes with replacing the VC and the EC with the EPC does not necessarily translate to code coverage. Moreover, the completeness of the model is crucial to observe the positive effects of the increased strength; if a graph does not model a key SUT behavior, no test criterion can cover that behavior. For this paper, we assume that the model sufficiently captures the relevant SUT behaviors, leaving the verification of the model as a separate topic.}

\subsection{Test oracles}

\added[id={I40}]{DJPlus only generates the executable tests for our experimental SUTs. The GraphWalker environment allows the automation of the test execution. The test oracles are implemented as assertions related to the elements of the graph model, where a test oracle indicates \emph{pass} if no assertion is violated and \emph{fail}, otherwise. In this work, the coverage of a test set is measured by counting the number of targeted model elements it covers. Therefore, our experimental results are not affected by the absence of a specific test oracle.}

% \subsection{Fairness of cross-criterion comparisons}
\subsection{Feasibility}

\added[]{To determine the feasibility of the test steps generated by DJPlus to achieve EPC, we compare them against the test steps generated by Graphwalker to achieve EC. This is not a like-for-like comparison because EPC is a stronger criterion and tend to require longer test suites. However, since Graphwalker test suites achieving EC for Testinium (the largest SUT in our study) are shown to be feasible \citep{Garousi+2021:JSS}, any smaller or equally large test suite achieving EC or stronger coverage can be said to be feasible as well.}

\section{Conclusions}\label{sec:conclusions}

In this paper, we proposed a novel method, DJPlus, that generates reduced amount of test steps within a minimized number of test cases for graph\hyp{}based models. We gave the reasons behind generating minimal test suites, demonstrated the suboptimalities \replaced[id={I12}]{with examples}{in previous methods}, and provided a detailed description of DJPlus. Then, we presented experimental results on four realistic SUTs within the popular GraphWalker environment. We showed that DJPlus generated the fewest test steps for the same coverage target, while the other methods generated 2 to 26 times more redundant test steps. We revealed that \added[id={I10}]{the }prime path criterion does not scale well, but \added[id={I10}]{the }edge\hyp{}pair criterion is feasible with DJPlus on our four experimental SUTs. Finally, we measured test execution times \added[id={I3}]{and performed an effect size analysis }to show that DJPlus-generated test suites indeed reduce test execution costs.

In the future, we seek to expand DJPlus to graphs with conditional edges, where the edge is enabled only if its condition holds. These conditions are called \emph{guards} and are already available as a feature in GraphWalker. We aim to modify DJPlus to identify the shortest detours that pass the guarded edges. Also, we plan to revise the test requirement generation method to obtain minimal test suites for coverage percentages below $100\%$. Finally, we will use mutation analysis to measure the fault detection capabilities of DJPlus\hyp{}generated tests.

\appendix

\noindent\deleted[id={I16}]{\textbf{Appendix A. Formal definitions}}

\section{Graph models in DJ}\label{sec:graphs}

\added[id={I9}]{In this section, we describe the necessary definitions for graph models to understand DJ and the motivating examples related to DJ / DJPlus. DJ, and its improved version, DJPlus, generate tests from graph models.}

\subsection{Graph models with simple graphs}

\added[id={I24}]{A \emph{graph model} $M = (G, v_e)$ in DJ is a pair of a \emph{simple graph} $\simplegraph{G}{V}{E}$ and an entry vertex $v_e \in V$, where $V$ is a set of vertices and $E \subseteq V^2$ is a set of edges connecting pairs of vertices. The model's graph is called \emph{simple} because it does not allow different edges connecting the same vertex pair.}

\deleted{A \emph{graph model} in MBT is a collection of triggerable SUT behaviors critical for testing purposes, e.g., application programming interface invocations, assertions, or executable code blocks, represented by vertices and connected by directed edges. When an edge connects one vertex to another, we refer to the vertices as the predecessor and the successor, respectively. In practice, an edge allows a test execution to trigger the successor's SUT behavior right after the predecessor's behavior. Conversely, behaviors not connected with an edge cannot follow one after the other.}

\deleted{Every graph model features an \emph{entry vertex}, representing the first SUT behavior triggered when the SUT starts execution. In reality, the SUT may feature multiple initial behaviors, depending on its inputs and the execution environment. However, this is not a problem from the MBT perspective. The graph model's entry vertex then represents a decision point between these initial behaviors corresponding to the successors of the entry vertex.}

\deleted{Every edge in a graph model is unique, meaning that there can be only one edge connecting a predecessor to a successor. Graph theory refers to graphs with unique edges as \emph{simple graphs}.}

\subsection{Test cases}\label{sec:graphs:testcases}

\added{A \emph{test case (test path)} $t$ is a vertex sequence whose first vertex is the entry vertex $t[0] = v_e$, and every consecutive vertex pair is an edge. The last vertex of a test case $t$ is its \emph{target}, defined as $\tgt{t} \defeq t[\len{t} - 1]$, where $\len{t}$ denotes the \emph{length} of the test case, calculated as the total number of its vertices. We also say that a test case $t$ contains $|t|$ many test steps, where every step corresponds to a vertex. We formally define a path as a subset of vertex sequences $P(G)$ that satisfy the edge property as}
\begin{equation*}
    \added{P(G) \defeq \setof{p}{p \in V^*, \forall i \in \N, 1 \leq i < \len{p} , (p[i-1], p[i]) \in E}.}
\end{equation*}

\subsection{Loops and cycles}

\added{A path $p$ is a \emph{loop} iff it has exactly two vertices and its vertices are the same, i.e., $|p| = 2, p[0] = \tgt{p}$. A \emph{cycle}, on the other hand, is a non-empty path whose target vertex has an edge to its first vertex, i.e., $\cycle{p} \defiff \len{p} \geq 1, (\tgt{p}, p[0]) \in E$. A simple graph $\simplegraph{G}{V}{E}$ is acyclic if it has no \emph{cycles}, i.e., $\ncycle{G} \defiff \forall p \in P(G), \ncycle{p}$.}

\deleted{Graph models permit \emph{loops}, i.e., an edge going from a vertex to itself, corresponding to the repetitive nature of a triggerable SUT behavior.}

\subsection{Sequence operations}\label{sec:graphs:seqops}

\added{Throughout the paper, we use several operations on vertex sequences. The first is taking the \emph{prefix (history)} of a sequence. The prefix of a vertex sequence $\ltrl{sq}$ is the subsequence of $\ltrl{sq}$ that excludes only the target vertex, i.e.,}
\begin{equation*}
    \pre{\ltrl{sq}} \defeq \begin{cases}
        \len{\ltrl{sq}} < 2 & () \\
        \len{\ltrl{sq}} \geq 2 & \ltrl{sq}\left[0 \ldots \len{\ltrl{sq}}{\shortminus}2\right]
    \end{cases}
\end{equation*}
\added{where $()$ is an empty sequence.}

\added{Rotating ($\symrot$) a vertex sequence $sq$ removes its target vertex and puts it in front, as in} 
\begin{equation*}
    \rot{\ltrl{sq}} \defeq \begin{cases}
        \len{\ltrl{sq}} = 0 & () \\
        \len{\ltrl{sq}} > 0 & \concat{(\tgt{\ltrl{sq}})}{\pre{\ltrl{sq}}}
    \end{cases}
\end{equation*}
\added{where $\symconcat$ denotes the concatenation of two sequences. In this paper, rotation involves cycles, which remain valid paths after any number of rotations.}

\added{Finally, the \emph{splice} of two paths $\splice{p}{q}$ is the set of shortest paths that start with $p[0]$ and end with $\tgt{q}$, as in}
\begin{equation*}
    \splice{p}{q} \defeq \setof{r}{r \in \ltrl{SP}, \forall s \in \ltrl{SP},  \len{r} \leq \len{s}}
\end{equation*}
where
\begin{equation*}
\ltrl{SP}\!\defeq\!\setof{r}{r \in P(G)\!\setminus\!V^0, \subpath{p}{r}, \subpath{q}{r}, r[0] = p[0], \tgt{r} = \tgt{q}}
\end{equation*}
\added{and $\sqsubseteq$ denotes the subpath relation between two paths.}

The literature leaves path splicing unnamed and denotes it with the set union symbol, like $p \cup q$~\citep{DwarakanathJankiti2014:ICTSS}.
For convenience, we named it and used the $\splice{p}{q}$ notation to avoid confusion with set union.
\deleted{Furthermore, the original authors define the splice not as a set but as an element, probably because splicing a pair of vertex, edge, edge\hyp{}pair, or prime path requirements always produces a unique path.
Our definition applies to all kinds of test requirements, even manually crafted ones.}

\subsection{Strong connectivity}

\added{We assume the reachability of every vertex $v \in V$ from the entry vertex $v_e$, and say that $v_e$ is strongly connected to every other vertex. We denote this fact as $\forall v \in V, v_e \leadsto v$ and define it as}
\begin{equation*}
    v_e \leadsto v \defiff \exists r, r \in \splice{(v_e)}{(v)}.
\end{equation*}

\deleted{We assume that every vertex is reachable from the entry vertex, so every SUT behavior is triggerable during test execution. In other words, the graph model excludes any unreachable SUT behavior.}

\subsection{Test suites, requirements, and criteria}

\added{A \emph{test suite} $T$ is a set of test cases. We define test requirements and criteria to measure the strength of a test suite.}

\deleted{A \emph{test suite} is a collection of test cases, e.g., $T = \lbrace t_1, t_2 \rbrace$. In MBT practice, developers or testers generate and execute test suites to ensure the software quality.}

\added{A test requirement $\ltrl{tr}$ is a path that must be covered by a test suite}\footnote{Broadly, a test requirement is a specific element of a software artifact that a test case must cover~\citep{AmmanOffut2008:CUP}. The definition a test requirement we use here is specific to graph models. The literature also uses testing target to refer to this concept~\cite{Arcuri+2012:TSE}.}\added{. So, the test suite must be generated such that every test requirement must be a subpath of at least one of the test cases.}

\deleted[id={I39}]{Thus, in graph models, a \emph{test requirement} is a non\hyp{}empty path whose final vertex represents a target behavior, which we refer to as the \emph{target vertex}.
Then, the vertices leading up to the target vertex form the \emph{history (prefix)} of the test requirement.
In practice, the SUT often uses hidden variables not included in the graph model, necessitating the testing of the same behavior with different histories.}

\added{A \emph{test criterion} is a set of test requirements. Formally, we define a test criterion $\ltrl{TR}$ as a subset of all possible test requirements $\ltrl{TR}(G)$ such that $\ltrl{TR}(G) \defeq P(G) \setminus V^0$, where we exclude empty paths, i.e., $V^0$.}

The literature defines general sets of test requirements known as graph coverage criteria~\citep[Part 2, Ch.~7]{AmmanOffut2008:CUP}. \added{A test requirement is \emph{redundant} iff it is a subpath of at least one of the other test requirements. A \emph{minimal} test criterion is free of redundant requirements. From now on, whenever we discuss a test criterion, we refer to a minimal one.}

\replaced{A criterion $\ltrl{TR}_1$ subsumes another criterion $\ltrl{TR}_2$ iff every requirement $\ltrl{tr}_2 \in \ltrl{TR}_2$ is a subpath of a $\ltrl{tr}_1 \in \ltrl{TR}_1$. Under this relation, we say that the criterion $\ltrl{TR}_1$ is stronger than $\ltrl{TR}_2$.}{If covering a set of test requirements ensures the coverage of every test requirement in another set, then the former set is \emph{stronger} than the latter.
The apparent tradeoff is that as the requirements get stronger, more test steps are necessary to achieve the same coverage percentage.} In this study, we focus on vertex, edge, edge\hyp{}pair, and prime path criteria\footnote{We refer to the literature for a comprehensive list of graph coverage criteria~\citep[Part 2, Ch.~7]{AmmanOffut2008:CUP}.}.

\replaced{\emph{The vertex criterion}}{\emph{Vertex criterion}} ($\crit{vc}$) \added{of a graph model} is the set of all test requirements with an empty history.
A test suite \replaced{that satisfies the vertex criterion}{with $100\%$ vertex coverage} \replaced{covers every vertex}{triggers every SUT behavior} at least once \deleted{during test execution}.
\deleted{However, vertex coverage is a sufficient test suite quality measure only if the histories do not matter at all, which is rarely the case in a real\hyp{}world SUT.}

Generating \replaced{each test requirement of}{the test requirements for} the vertex criterion is str\-aight\-forward; \deleted{we} take every vertex of the graph as a separate test requirement.

\replaced{\emph{The edge criterion}}{\emph{Edge criterion}} ($\crit{ec}$) is the set of all test requirements with a history up to length one.
A test suite \replaced{that satisfies the edge criterion}{with $100\%$ edge coverage} reaches a target \replaced{vertex}{behavior} from every possible predecessor.
\replaced{T}{Naturally, t}he edge criterion is stronger than the vertex criterion.

To get the test requirements for the edge criterion, one must take every edge of the graph as a separate test requirement.
\deleted{Although the definition includes requirements with empty histories (vertices), those requirements are redundant.}

\replaced{\emph{The edge\hyp{}pair criterion}}{\emph{Edge\hyp{}pair criterion}} ($\crit{epc}$) is the set of all test requirements with a history up to length two.
A straightforward way to generate these test requirements is to compute all non\hyp{}empty paths up to length three and then remove the redundant ones\added{, i.e., the paths that are strict subpaths of others}.
The edge\hyp{}pair criterion is stronger than both edge and vertex criteria.

\replaced{\emph{The prime path criterion}}{\emph{Prime path criterion}} ($\crit{ppc}$) is the set of all test requirements with a unique history of arbitrary length, where a unique history contains no vertex more than once, and can include the target vertex only at the beginning.
The prime path criterion is stronger than vertex and edge criteria, but not edge\hyp{}pair criterion.

\replaced{As a final note, a}{A}lgorithms that generate a minimal set of requirements for \replaced{the} prime path criterion are well\hyp{}known (see \ref{sec:computations:ppc}).

\deleted{Finally, to measure coverage, we calculate the percentage of test requirements that a test suite covers.}

\section{Known algorithms}\label{sec:computations}
To make this paper self\hyp{}contained, we present the known algorithms we used in this section.

\subsection{Computing a splice}
\label{sec:computations:splice}

\begin{elsalgorithm}
\caption{Path splicing ($\spliceG{p}{q}{G}$)}\label{alg:splice}
\KwIn{\\\Indp%
    $\simplegraph{G}{V}{E}$ \\%
    $p \in P(G) \setminus V^0$ \\%
    $q \in P(G) \setminus V^0$%
}
\KwOut{\\\Indp%
$R \subseteq P(G) \setminus V^0$%
}
$i \leftarrow \min(\len{p},\len{q})$\;\label{alg:splice:overlap:begin}
\lWhile{$i > 0$ \And $p[\len{p}{\shortminus}i \ldots \len{p}{\shortminus}1] \!\neq\! q[0 \ldots i{\shortminus}1]$}{
    $i \leftarrow i - 1$%\;
}\label{alg:splice:overlap:end}
\lIf{i > 0}{
    $R \leftarrow \left\lbrace \concat{p}{q[i \ldots \len{q}{\shortminus}1]} \right\rbrace $\label{alg:splice:concat}
}\lElse{%
    $R \leftarrow \lbrace \rbrace$ \nForEach $c \in C_G(p, q)$ \nDo $R \leftarrow R \cup \lbrace \concat{p}{\concat{c}{q}} \rbrace$
}
\end{elsalgorithm}
 
We present the steps to splice two paths in \cref{alg:splice}. In lines~\ref{alg:splice:overlap:begin}-\ref{alg:splice:overlap:end}, we compute the length of overlap between $p$ and $q$, i.e., the last $i$ vertices of $p$ that are equal to the first $i$ vertices of $q$.
If there is an overlap ($i > 0$), the only splice is the concatenation of $p$ with the remainder of $q$, as in line~\ref{alg:splice:concat}.
Else, we concatenate $p$ with $c$ and then $q$, where $c \in C_G(p, q)$ is every shortest path that connects $p$ and $q$.
Computing $c$ is equivalent to computing all shortest paths from \replaced{$\tgt{p}$}{the last vertex of $p$} to \replaced{$q[0]$}{the first vertex of $q$}.
Another option is to compute all\hyp{}pairs shortest paths beforehand, which is useful from the MBT perspective.
We refer to the literature for the ever\hyp{}growing research to accomplish these tasks~\citep{Thorup2004:JCSS:STOC, Williams2018:SiamComp}.
Finally, if a path that connects $p$ to $q$ does not exist, then the splice remains an empty set.

The original authors of path splicing did not consider the fact that there could be multiple splices of two vertices, probably because the test requirement pairs from vertex, edge, edge\hyp{}pair, and prime path coverages all generate at most one splice~\citep{DwarakanathJankiti2014:ICTSS}.
Our variant of path splicing enables the test generation to work with any set of test requirements, even manually crafted ones.

\subsection{Computing prime paths}
\label{sec:computations:ppc}

\begin{elsalgorithm}
\caption{Prime path generator}\label{alg:ppc}
\KwIn{\\\Indp%
    $\simplegraph{G}{V}{E}$ \\%
}
\KwOut{\\\Indp%
    $\crit{ppc} \subseteq \ltrl{TR}(G)$%
}
$\crit{ppc} \leftarrow E$ \;
\Repeat{$c = 0$}{
    $c \leftarrow 0, U \leftarrow \crit{ppc}$ \;
    \ForEach{$\ltrl{tr}_1 \in U$ \St $\ncycle{\ltrl{tr}_1}$}{
        \ForEach{$(v, \ltrl{tr}_1[0]) \in E$ \St $v \notin \pre{\ltrl{tr}_1}$}{
            $c \leftarrow 1$\;
            \ForEach{$\ltrl{tr}_2 \in \crit{ppc}$ \St $\subpath{\ltrl{tr}_2}{\concat{(v)}{\ltrl{tr}_1}}$}{
                $\crit{ppc} \leftarrow \crit{ppc} \setminus \ltrl{tr}_2$\;
            }
            $\crit{ppc} \leftarrow \crit{ppc} \cup \lbrace \concat{(v)}{\ltrl{tr}_1} \rbrace $\;
        }
    }
}
\end{elsalgorithm}

Algorithms that compute a minimal set of test requirements for prime paths are well\hyp{}known~\citep[Part 2, Ch.~7]{AmmanOffut2008:CUP}. The literature also features equivalent variants of the same algorithm~\citep{DwarakanathJankiti2014:ICTSS}. We provide our variant in \cref{alg:ppc}, which follows our definition of the prime path criterion closely.

\subsection{Computing splice graphs}
\label{sec:computations:splicegraph}

\begin{elsalgorithm}
    \caption{Splice graph generation}\label{alg:splicegraph:formal}
    \KwIn{\\\Indp%
        $\graphmodel{M}{G}{V}{E}{v_e}$\\%
        $\ltrl{TR} \subseteq \ltrl{TR}(G)$
    }
    \KwOut{\\\Indp%
        $\hypergraph{S}{G}{V_S}{E_S}$
    }
    $V_S \leftarrow \lbrace (v_e) \rbrace \cup \ltrl{TR}, E_S \leftarrow \lbrace \rbrace$ \;
    \ForEach{$v \in V_S$}{
        \ForEach{$y \in \ltrl{TR} \setminus \lbrace v \rbrace$}{
            $E_S \leftarrow E_S \cup \lbrace (v, y) \rbrace$\;
            \ForEach{$r \in \spliceG{v}{y}{G}$ \And \Whl $(v, y) \in E_S$}{\label{alg:splicegraph:formal:splice}
                \lIf{$\exists z, z \in \ltrl{TR} \!\setminus\! \lbrace v, y \rbrace, \subpath{z}{r}$}{
                    $E_S \leftarrow E_S \!\setminus\! \lbrace (v, y) \rbrace$
                }
            }
        }
    }
    $S \leftarrow \left(V_S, E_S\right)$\;
\end{elsalgorithm}

We present the formal steps to generate a splice graph in \cref{alg:splicegraph:formal}.
We modify the original algorithm to cover the case when there are more than one splice $\spliceG{v}{y}{G}$.
For more details, we refer to the literature that first presented this algorithm \citep[Algorithm 2]{DwarakanathJankiti2014:ICTSS}.

\subsection{Computing hypergraphs}
\label{sec:computations:hypergraph}

\begin{elsalgorithm}
    \caption{Hypergraph generation}\label{alg:hypergraph:formal}
    \KwIn{\\\Indp%
        $\simplegraph{G}{V}{E}$ \\
        $\hypergraph{S}{G}{V_S}{E_S}$
    }
    \KwOut{\\\Indp%
        $H \in (S \cup H(S) \cup H(H(S)) \cup \ldots)^*$ \St $H[0] = S, \ncycle{\tgt{H}}, \forall i [i \in N, 0 < i < \len{H}] \rightarrow \hypergraph{H[i]}{H[i{\shortminus}1]}{V_{H[i]}}{E_{H[i]}}$
    }
    $H \leftarrow (S), i \leftarrow 0$ \;
    \Repeat{$H[i{\shortminus}1] = H[i]$}{
        $i \!\leftarrow\! i + 1, H[i] \!\leftarrow\! H[i{\shortminus}1], s \!\leftarrow\! ()$ \lFor{$e \in E_{H[i]}$}{$s \!\leftarrow\! \concat{s}{(e)}$}
        \While{$\len{s} > 0$ \And $H[i{\shortminus}1] = H[i]$}{
            $p \leftarrow \tgt{s}, s \leftarrow \pre{s}$ \;
            \uIf{$\cycle{p}$}{
                $V_{H[i]} \leftarrow V_{H[i]} \cup \lbrace (h_i) \rbrace$ \;
                \ForEach{$v \in V_{H[i]} \setminus \lbrace (h_i) \rbrace$ \St $v \notin p$}{
                    \If{$\exists j, j \in \N, j < \len{p}, (v, p[j]) \in E_{H[i]}$}{
                        $E_{H[i]} \leftarrow E_{H[i]} \setminus (v, p[j])$\;
                        $E_{H[i]} \leftarrow E_{H[i]} \cup (v, h_i)$\;
                    }
                    \If{$\exists j, j \in \N, j < \len{p}, (p[j], v) \in E_{H[i]}$}{
                        $E_{H[i]} \leftarrow E_{H[i]} \setminus (p[j], v)$\;
                        $E_{H[i]} \leftarrow E_{H[i]} \cup (h_i, v)$\;
                    }
                }
                \lForEach{$v \in p$}{$V_{H[i]} \leftarrow V_{H[i]} \setminus \lbrace v \rbrace$}
            }\Else{
                \ForEach{$v \!\in\! V_{H[i]}$ \St $v \!\notin\! \pre{p}, (v, p[0]) \!\in\! E_{H[i]}$}{
                    $s \leftarrow \concat{s}{(\concat{(v)}{p})}$ \;
                }
            }
        }
        $H \leftarrow \concat{H}{(H[i])}$\;
    }
    $H \leftarrow \pre{H}$
\end{elsalgorithm}

We present the formal steps to generate hypergraphs in \cref{alg:hypergraph:formal}.
\added[id={I45}]{}Notice that in this variant, we chose to remove the \emph{shortest cycle} first because finding the shortest cycle is computationally cheaper and relatively easier to implement than a longest\hyp{}simple\hyp{}cycle\hyp{}first heuristic.
We refer to the literature for an algorithm that finds the longest simple cycle in a graph~\citep{Anaqreh+2023:arXiv}.

\deleted{Unlike its informal counterpart, this algorithm, outputs a sequence of hyperpaths, replacing one cycle at a time. The reason behind this sequence is to remember the edges of the cycles in the unwinding phase. In the informal descriptions, we kept this information implicit.}

\subsection{Unwinding}
\label{sec:computations:unwinding}

\begin{elsalgorithm}
    \caption{Unwinding hyperpaths}\label{alg:unwinding:formal}
    \KwIn{\\\Indp%
        $\simplegraph{H[0]}{V_{H[0]}}{E_{H[0]}}$ \\
        $\forall i [i \in \N, 0 < i < \len{H}] \rightarrow \hypergraph{H[i]}{H[i{\shortminus}1]}{V_{H[i]}}{E_{H[i]}}$\\
        $\pi \in V^{**}_{\tgt{H}}$
    }
    \KwOut{\\\Indp%
        $p \in P(H[0])^*$
    }
    \For{$i = \len{H} - 1$ \To $1$}{
        $p \leftarrow \pi$ \;
        \For{$j = 0$ \To $\len{\pi} - 1$}{
            $\pi[j] \leftarrow ()$\;
            \For{$k = 0$ \To $\len{\pi[j]} - 1$}{
                \lIf{$p[j][k] \in V_{H[i{\shortminus}1]}$}{
                    $\pi[j] \leftarrow \concat{\pi[j]}{(p[j][k])}$\label{alg:unwinding:formal:straightconcat}
                }\Else{
                    $q \leftarrow (v)$ \St $v \in V_{H[i{\shortminus}1]} \setminus V_{H[i]}$\;\label{alg:unwinding:formal:reconstruct:begin}
                    \While{$\ncycle{q}$}{
                        $q \leftarrow \concat{(v)}{q}$ \St $v \!\in\! V_{H[i{\shortminus}1]} \!\setminus\! V_{H[i]}, (v, q[0]) \!\in\! E_{H[i{\shortminus}1]}$\;
                    }\label{alg:unwinding:formal:reconstruct:end}
                    $r \leftarrow \pre{q}, s_1 \leftarrow s$ \St $s \in \spliceG{\spliceG{\pi[j]}{r}{H[i-1]}}{p[j][k+1\ldots\len{p[j]}{\shortminus}1]}{H[i-1]}$\;\label{alg:unwinding:formal:rotate:begin}
                    \nRepeat{$\len{r} - 1$}{
                        $r \leftarrow \rot{r}$\;    
                        $s_2 \leftarrow s$ \St $s \in \spliceG{\spliceG{\pi[j]}{r}{H[i-1]}}{p[j][k+1\ldots\len{p[j]}{\shortminus}1]}{H[i-1]}$\;
                        \lIf {$\len{s_2} < \len{s_1}$}{$s_1 \leftarrow s_2$}
                    }\label{alg:unwinding:formal:rotate:end}
                    $\pi[j] \leftarrow \pre{s_1}$\;\label{alg:unwinding:formal:spliceconcat}
                }
            }
        }
    }
\end{elsalgorithm}

We present the formal steps of unwinding hyperpaths in \cref{alg:unwinding:formal}.
This algorithm takes a sequence of hypergraphs $H$, and a sequence of vertex sequences of the final hypergraph, e.g., \cref{eq:exhyperseqreduced}.
It outputs a sequence of paths of the first hypergraph.

\cref{alg:unwinding:formal} converts the vertex sequences of an hypergraph to vertex sequences of the preceding hypergraph until there is no preceding hypergraph.
In line~\ref{alg:unwinding:formal:straightconcat}, it copies the elements that are in both hypergraphs.
Otherwise, it reconstructs the cycle represented by the hyperpath in lines~\ref{alg:unwinding:formal:reconstruct:begin}-\ref{alg:unwinding:formal:reconstruct:end}.
Then, finds the rotation of this cycle's prefix that generates the shortest splice in lines~\ref{alg:unwinding:formal:rotate:begin}-\ref{alg:unwinding:formal:rotate:end}.
This shortest splice replaces the path in line~\ref{alg:unwinding:formal:spliceconcat}.

\subsection{Unrolling}
\label{sec:computations:unrolling}

\begin{elsalgorithm}
    \caption{Unrolling hyperpaths}\label{alg:unrolling:formal}
    \KwIn{\\\Indp%
        $\simplegraph{H[0]}{V_{H[0]}}{E_{H[0]}}$ \\
        $\forall i [i \in \N, i < \len{H}] \rightarrow \hypergraph{H[i]}{H[i{\shortminus}1]}{V_{H[i]}}{E_{H[i]}}, \ncycle{\tgt{H}}$
    }
    \KwOut{\\\Indp%
        $\simplegraph{\ltrl{RFG}}{V_{\ltrl{RFG}}}{E_{\ltrl{RFG}}}, \ncycle{\ltrl{RFG}}$
    }
    $V_{\ltrl{RFG}} \leftarrow V_{\tgt{H}}, E_{\ltrl{RFG}} \leftarrow E_{\tgt{H}}$\label{alg:unrolling:formal:1}\;
    $\ltrl{RFG} \leftarrow \text{minimize}(V_{\ltrl{RFG}}, E_{\ltrl{RFG}})$\label{alg:unrolling:formal:2}\;
    \For{$i = \len{H} - 1$ \To $1$\label{alg:unrolling:formal:3}}{
        $h \sim h[0\ldots\len{h}], h' \sim h'[0\ldots\len{h}]$\label{alg:unrolling:formal:id} \St $h \in V_{H[i]} \!\setminus\! V_{H[i{\shortminus}1]}, \forall \! j [j \in \N, j < \len{h}] \rightarrow h[j] \in V_{H[i{\shortminus}1]} \!\setminus\! V_{H[i]}$\; 
        \For{$j = 0$ \To $\len{h} - 1$}{
            $V_{\ltrl{RFG}} \leftarrow V_{\ltrl{RFG}} \cup \lbrace h[j], h'[j] \rbrace$\label{alg:unrolling:formal:vadd}\;
            $E_{\ltrl{RFG}} \leftarrow E_{\ltrl{RFG}} \cup \lbrace (h[j], ()), (h'[j], ()) \rbrace$\label{alg:unrolling:formal:esink}\;
            \ForEach{$v \in V_{H[i{\shortminus}1]} \!\cap\! V_{H[i]}$ \St $(v, h[j]) \in E_{H[i{\shortminus}1]}$}{
                $E_{\ltrl{RFG}} \leftarrow E_{\ltrl{RFG}} \cup \lbrace (v, h[j]), (v, h'[j]) \rbrace$\label{alg:unrolling:formal:efrom}\;
            }
            \ForEach{$v \in V_{H[i{\shortminus}1]} \!\cap\! V_{H[i]}$ \St $(h[j], v) \in E_{H[i{\shortminus}1]}$}{
                $E_{\ltrl{RFG}} \leftarrow E_{\ltrl{RFG}} \cup \lbrace (h[j], v) \rbrace$\label{alg:unrolling:formal:eto1}\;
                \lIf{$(v, h') \notin E_{\ltrl{RFG}}$}{$E_{\ltrl{RFG}} \leftarrow E_{\ltrl{RFG}} \cup \lbrace (h'[j], v) \rbrace$\label{alg:unrolling:formal:eto2}}
            }
            \If{$j > 0$}{
                $E_{\ltrl{RFG}} \leftarrow E_{\ltrl{RFG}} \cup \lbrace (h[j{\shortminus}1], h[j]), (h'[j{\shortminus}1], h'[j]) \rbrace$\label{alg:unrolling:formal:eadd}\;
            }
        }
        $E_{\ltrl{RFG}} \leftarrow E_{\ltrl{RFG}} \cup \lbrace (\tgt{h}, h'[0]) \rbrace$\label{alg:unrolling:formal:elast}\;
        \ForEach{$v \in V_{\ltrl{RFG}}$}{
            $E_{\ltrl{RFG}} \leftarrow E_{\ltrl{RFG}} \setminus \lbrace (h, v), (v, h) \rbrace$\label{alg:unrolling:formal:eremove}\;
        }
        $V_{\ltrl{RFG}} \leftarrow V_{\ltrl{RFG}} \setminus \lbrace h \rbrace$\label{alg:unrolling:formal:vremove}\;
        $\ltrl{RFG} \leftarrow \text{minimize}(V_{\ltrl{RFG}}, E_{\ltrl{RFG}})$\label{alg:unrolling:formal:lastmin}\;
    }
\end{elsalgorithm}

We present the formal steps of our novel unrolling algorithm in \cref{alg:unrolling:formal}.
The input is a sequence of hypergraphs, starting from $H[0]$.
Each successor hypergraph's vertices are some paths of the predecessor.
The final hypergraph $\tgt{H}$ is acyclic.
The output ($\ltrl{RFG}$) is an acyclic simple graph.

\cref{alg:unrolling:formal} starts by copying the final hypergraph $\tgt{H}$ to $\ltrl{RFG}$ in line~\ref{alg:unrolling:formal:1}. 
$tgt(H)$ is acyclic, so the algorithm performs flow minimization on $\ltrl{RFG}$ in line~\ref{alg:unrolling:formal:2}, which removes some of the edges.

In line~\ref{alg:unrolling:formal:3}, \cref{alg:unrolling:formal} loops through all hyperpaths in the reverse order of their construction.
At every iteration, it reconstructs the vertices of the hyperpath $h \in V_{H[i]} \setminus V_{H[i{\shortminus}1]}$ in line~\ref{alg:unrolling:formal:id}.
There is only one such hyperpath because \cref{alg:hypergraph:formal} adds one hyperpath per new hypergraph.
The reconstruction is arbitrary, e.g., $h = (h[0], h[1], h[2])$ could be reconstructed as $h = (h[2], h[0], h[1])$ because it corresponds to a cycle of the preceding hypergraph $H[i-1]$.
The arbitrariness is irrelevant because the order of the vertices is not important.
For every reconstructed vertex $h[j]$, the algorithm also constructs $h'[j]$.

\cref{alg:unrolling:formal} adds every vertex of $h$'s cycle to $\ltrl{RFG}$ in line~\ref{alg:unrolling:formal:vadd}.
Note that it adds every vertex twice, like $h[j]$ and $h'[j]$.
Then, it connects every added vertex to the sink in line~\ref{alg:unrolling:formal:esink}.

In line~\ref{alg:unrolling:formal:efrom}, the algorithm reconstructs the hyperpath's incoming edges.
In lines~\ref{alg:unrolling:formal:eto1} and \ref{alg:unrolling:formal:eto2}, the algorithm reconstructs the hyperpath's outgoing edges, where the if statement preserves the acyclicity of $\ltrl{RFG}$, e.g., in \cref{fig:unrollBh1}, $((v_1), (v_2))$ is an edge, so technically, $((v_1)', (v_2))$ is also valid.
Thanks to this if statement, the algorithm ignores that edge to avoid introducing cycles.

Every consecutive vertex pair in the cycle forms a valid edge, so the algorithm makes these connections in line~\ref{alg:unrolling:formal:eadd}.
Finally, it connects the last vertex of the cycle $\tgt{h}$ to $h'[0]$ in line~\ref{alg:unrolling:formal:elast}.
Notice that the algorithm refrains from connecting $(\tgt{h}, h[0])$, $(\tgt{h'}, h[0])$, and $(\tgt{h'}, h'[0])$, all of which would introduce a cycle.
As a result, the algorithm preserves the acyclicity of $\ltrl{RFG}$.

\cref{alg:unrolling:formal} removes all the edges to and from $h$ in line~\ref{alg:unrolling:formal:eremove} and finally removes $h$ in line~\ref{alg:unrolling:formal:vremove}. It does not remove $h'$ because that is the only way to traverse the same vertex thrice or more times in the test suite, whenever it is necessary.
As the resulting $\ltrl{RFG}$ remains acyclic, the algorithm performs a re\hyp{}minimization in line~\ref{alg:unrolling:formal:lastmin}, where every vertex decorated with an apostrophe must have at least zero outgoing flow.
Every other vertex must have at least one outgoing flow.
Thus, re\hyp{}minimization removes some edges and some of the decorated vertices.
In the end, the resulting $\ltrl{RFG}$ contains only test requirements along with the source and sink vertices.

\noindent\deleted[id={I27}]{\textbf{Appendix C. Detailed test generation results}}

\deleted{We present the average number of test steps generated by Random, QRandom, and DJPlus for $50\%$, $60\%$, $70\%$, $80\%$, $90\%$, and $100\%$ vertex and edge coverages on all four experimental SUTs with the $30$ experimental seed values in Fig. C. 1.
We compare the number of test steps of the three test generation methods (Random, QRandom, and DJPlus) for each SUT/target criterion/target percentage triple, e.g., Testinium/EC/100.
Thus, we make comparisons in $4 \times 6 \times 2 = 48$ distinct experimental configurations.}

\deleted{According to Fig. C. 1, we observe that DJPlus generated the smallest test suite in $38$ of the $48$ configurations.
The configurations QRandom generates a smaller test suite for are mostly when the model is small, e.g., TLC/EC/100, or when the target percentage is low, e.g., Parabank/EC/50.
Random never generated a minimal test suite.}

\deleted{Since DJPlus generated the smallest test suite in most experimental configurations, we conclude that GraphWalker must be generating test steps that are redundant for the coverage target.
However, we cannot quantify the amount of redundancy directly from Fig. C. 1. 
Thus, we propose a redundancy measure in ?.}

\deleted{From the results in Fig. C. 1, we observe that our strategy of generating test cases for $100\%$ coverage and trimming them to adapt to a lower coverage target is highly inefficient.
In the future, we aim to evaluate tests generated for an automatically reduced set of test requirements and see if that mitigates some of the inefficiency.}

\section{Computing redundancies}\label{sec:redundancies}
To calculate the averages in \cref{tab:redundancies}, we start by rewriting \cref{eq:epcrf} as
\begin{equation*}
    \rf(T) = \rf(\tau) = \frac{\dlen{\tau}-2\len{\tau}}{\cov{epc}(\tau)\len{\crit{epc}}} - 1\text{, where}
\end{equation*}
\begin{enumerate}
    \item $\tau = \setof{t}{t \in T, \len{t} > 2}$,
    \item $\dlen{\tau} = \sum_{t \in \tau}\len{t}$, and
    \item $\cov{epc}(T) = \cov{epc}(\tau)$; as no test case $t \in T \setminus \tau$ has edge pairs.
\end{enumerate}

The redundancy factor is undefined for empty test suites and graphs with no edge pairs.
Thankfully, $\tau$ is never empty and all graphs have some edge pairs in our experiments.

We compute the relative redundancy of $\tau$ over $U$, where $U$ corresponds to the test suite with the same experimental settings but generated by DJPlus. The relative redundancy $\rr(\tau,U)$ is undefined when $\rf(U) = 0$; we compare relative redundancies only when DJPlus also has some redundancy.

\begin{elstable}
    \caption{Redundancy factors and relative redundancies of two test suites for Testinium.}
    \label{tab:exredundancy}
    \resizebox{\columnwidth}{!}{
        \begin{tabular}{rrrrrrr}
            \hline
             Test suite ($\tau$) & $\dlen{\tau}$ & $\len{\tau}$ & $\cov{epc}(\tau)$ & $\len{\crit{epc}}$ & $\rf(\tau)$ & $\rr(\tau, U)$ \\
             \hline
             DJPLUS/VC\deleted[id={I27}]{/100} & $285$ & $1$ & $0.25$ & $\num{1033}$ & $0.0958$ & $-$\\
             QRND/VC\deleted[id={I27}]{/100}/16 & $537$ & $1$ & $0.35$ & $\num{1033}$ & $0.0797$ & $4.006$x \\
             \hline
             \multicolumn{7}{r}{\added[id={I13}]{$\dlen{\tau}$: Total number of test steps in $\tau$.}}\\
             \multicolumn{7}{r}{\added[id={I13}]{$\len{\tau}$: Total number of test cases in $\tau$.}}
        \end{tabular}
    }
    % \resizebox{\columnwidth}{!}{
    %     \pgfplotstabletypeset[
    %         column type={r},
    %         columns/Configuration/.style={string type,column name={Test suite ($\tau$)}},
    %         columns/dlenT/.style={column name={$\dlen{T}$}},
    %         columns/lenT/.style={column name={$\len{T}$}},
    %         columns/dlenPi/.style={column name={$\dlen{\tau}$}},
    %         columns/lenPi/.style={column name={$\len{\tau}$}},
    %         columns/{C/EPC}/.style={column name={$\cov{epc}(\tau)$},fixed zerofill,precision=2},
    %         columns/{TR/EPC}/.style={column name={$\len{\crit{epc}}$}},
    %         columns/RF/.style={column name={$\rf(\tau)$},fixed zerofill,precision=4},
    %         columns/RR/.style={column name={$\rr(\tau,U)$},fixed zerofill,precision=1,
    %             postproc cell content/.append style={
    %                 /pgfplots/table/@cell content/.add={}{\if##1\empty\else\%\fi},
    %             }
    %         }%,
    %         % every last row/.style={
    %         %     after row={
    %         %         \bottomrule
    %         %         \multicolumn{7}{r}{$\dlen{\tau}$: Total number of test steps in $\tau$.} \\
    %         %         \multicolumn{7}{r}{$\len{\tau}$: Total number of test cases in $\tau$.}
    %         %     }
    %         % }
    %     ]{tables-exredundancy.csv}
    % }
\end{elstable}

The two rows in \cref{tab:exredundancy} illustrate the computation of the redundancy factor and the relative redundancy for two arbitrary experimental configurations in Testinium.
The redundancy factor for the test suite generated by DJPlus for \replaced[id={I27}]{the vertex criterion (DJPlus/VC)}{$100\%$ vertex coverage (DJPlus/VC/100)} is
\begin{equation*}
    \rf(\tau) = \frac{285 - 2(1)}{(0.25)(\num{1033})} - 1 \approx 0.0958 > 0
\end{equation*}
which means that $\tau$ covers some edge pairs more than once.
For this configuration, $\tau = U$, so the relative redundancy is
\begin{equation*}
    \rr(\tau,U) = \deleted[id={I30}]{100}\left(\frac{\rf(\tau)}{\rf(U)} - 1\right) = \deleted{100}\left(\frac{\rf(U)}{\rf(U)} - 1\right) = 0\%
\end{equation*}
which is meaningless, so we do not report it.
The relative redundancy is useful for test suites generated by other methods.

The second test suite comes from QRandom, targeting \deleted[id={I27}]{$100\%$} vertex coverage, using the $16^{\text{th}}$ experimental seed value.
This configuration results in a higher redundancy factor, meaning that $\tau$ has more redundancies than the DJPlus\hyp{}generated test suite.
We calculate its relative redundancy as
\begin{equation*}
    \rr(\tau, U) = \deleted[id={I30}]{100}\frac{0.4797}{0.0958} - 1 \approx 4.006
\end{equation*}
which means that QRandom was \replaced[id={I30}]{$4$ times}{$400\%$} more redundant than DJPlus for this configuration, while \replaced{getting}{managing} only $10\%$ more edge\hyp{}pair coverage for this cost.

Overall, we calculate the average across all vertex and edge coverage percentages, all SUTs, and all seed values to obtain one redundancy factor and one relative redundancy percentage per test generation method. 

\noindent\deleted[id={I60}]{\textbf{Appendix D. Proof of critical edge pair rarity}}%\label{sec:criticalproof}

\deleted{In this section, we provide an informal proof that a critical edge pair is rarer than a critical vertex or edge in a graph model.}

\deleted{Consider a simple graph with one critical edge pair.
Every vertex and edge of this pair is also critical, i.e., the graph contains at least three vertices and two edges, all of which are critical.
On the other hand, the existence of a critical vertex or edge does not imply the existence of a critical edge pair.
Thus, critical edge pairs are less likely to occur than critical vertices and edges.}

\section{Line graphs}
\label{sec:linegraphs}

\added[id={I50}]{In this section, we formally define line graphs and show the relationship between test criteria over a graph and its line graph.}

\added{First formally define a multigraph whose line graph is utilized in \cref{sec:method:implementation}. A multigraph $\ltrl{MG}$ is a triple such that $\multigraph{\ltrl{MG}}{V}{\ltrl{EL}}{E}$, where $\ltrl{EL}$ is a set of edge labels that enable multiple edges between the same vertex pair in the set of edges $E$.}

A \emph{line graph} $L(\ltrl{MG})$ of a multigraph $\multigraph{\ltrl{MG}}{V}{\ltrl{EL}}{E}$ is a simple graph $\simplegraph{L(\ltrl{MG})}{V_L}{E_L}$, such that
\begin{enumerate}
    \item $V_L = \ltrl{EL}$ and
    \item $\exists (v_0, a, v_1) \in E, \exists (v_1, b, v_2) \in E \leftrightarrow (a, b) \in E_L$.
\end{enumerate}

\added{We prove the equivalence between the VC of $L(\ltrl{MG})$ and the EC of $\ltrl{MG}$ by using the fact that $V_L = \ltrl{EL}$. From this fact, it follows that covering all the vertices in the line graph must cover all the edge labels in the multigraph, and vice versa.}

\added{We prove the equivalence between the EC of $L(\ltrl{MG})$ and the EPC of $\ltrl{MG}$ by considering the set of all edge pairs in $\ltrl{MG}$, which can be shown as $\ltrl{EL}^2$. Since $V_L = \ltrl{EL}$, the set of all vertex pairs in the line graph $V_L^2$ must be the same as $\ltrl{EL}^2$. Therefore, the EC of the line graph is equivalent to the EPC of the multigraph.}

\section*{CRediT authorship contribution statement}\label{sec:CRediTauthorshipcontributionstatement}

\credit{Yavuz K{\"o}ro\u{g}lu}{Conceptualization, Methodology, Software, Investigation, Data curation, Formal analysis, Writing -- original draft, Writing -- review \& editing, Funding acquisition, Visualization}
\credit{Mutlu Beyaz{\i}t}{Conceptualization, Validation, Investigation, Resources, Writing -- original draft, Writing -- review \& editing}
\credit{Onur K{\i}l{\i}n\c{c}\c{c}eker}{Conceptualization, Validation, Investigation, Resources, Writing -- original draft, Writing -- review \& editing}
\credit{Serge Demeyer}{Writing -- review \& editing, Supervision, Funding acquisition, Project administration}
\credit{Franz Wotawa}{Writing -- review \& editing, Supervision, Funding acquisition, Project administration}

%\insertcredits
\insertcreditsstatement

\section*{Declaration of competing interest}\label{sec:declarationofcompetinginterest}

The authors declare that they have no known competing financial interests or personal relationships that could have appeared to influence the work reported in this paper.

\section*{Acknowledgments}\label{sec:acknowledgments}

The work described in this paper is supported by
(a) the {\.I}stanbul Technical University Scientific Research Project (BAP) under identification number 47709,
(b) the Austrian Science Fund (FWF) Cluster of Excellence Bilateral AI under contract number 10.55776/COE12,
(c) the European Union under Horizon Europe via the ``INNO2MARE'' project under grant number 101087348,
(d) the Research Foundation Flanders (FWO) via the project ``Basecamp Zero'' under grant number S000323N, and 
(e) the Agency for Innovation \& Entrepreneurship (VLAIO) via the project ``TTRUST'' under grant number HBC20230612.

\section*{Data availability}\label{sec:dataavailability}

The dataset \added[id={I8}]{and the replication package} \replaced{are}{is} available at \url{https://doi.org/10.5281/zenodo.15556723} (Version v2). \added{This package includes instructions on how to replicate our experiments.}

GWPlus is available at \url{https://zenodo.org/records/15556723/files/gwplus.zip?download=1}\added{, so it can be used on other models}.

\bibliographystyle{elsarticle-harv}
\bibliography{rev1citations}

@misc{AhoLee:1987,
  title        = {Efficient algorithms for constructing testing sets, covering paths, and minimum flows},
  author       = {Aho, Alfred V and Lee, David},
  howpublished = {AT\&T Bell Laboratories Tech. Memo. CSTR159, pp. 1--15},
  pages        = {1--15},
  year         = {1987},
  url          = {https://tuhs.v6sh.org/UnixArchiveMirror/Documentation/TechReports/Bell_Labs/CSTRs/159.pdf}
}

@article{Alegroth+2022:ESE,
  author       = {Emil Al{\'{e}}groth and
                  Kristian Karl and
                  Helena Rosshagen and
                  Tomas Helmfridsson and
                  Nils Olsson},
  title        = {Practitioners' best practices to Adopt, Use or Abandon Model-based
                  Testing with Graphical models for Software-intensive Systems},
  journal      = {Empir. Softw. Eng.},
  volume       = {27},
  number       = {5},
  pages        = {103},
  year         = {2022},
  doi          = {10.1007/S10664-022-10145-2},
  bibsource    = {dblp computer science bibliography, https://dblp.org}
}

@book{AmmanOffut2008:CUP,
  author       = {Paul Ammann and
                  Jeff Offutt},
  title        = {Introduction to Software Testing},
  publisher    = {Cambridge University Press},
  year         = {2016},
  doi          = {10.1017/9781316771273},
  isbn         = {9781107172012},
  bibsource    = {dblp computer science bibliography, https://dblp.org}
}

@article{Anaqreh+2023:arXiv,
  author       = {Ahmad T. Anaqreh and
                  Bogl{\'{a}}rka G.{-}T{\'{o}}th and
                  Tam{\'{a}}s Vink{\'{o}}},
  title        = {Exact Methods for the Longest Induced Cycle Problem},
  journal      = {CoRR},
  volume       = {abs/2311.15899},
  year         = {2023},
  doi          = {10.48550/ARXIV.2311.15899},
  eprinttype   = {arXiv},
  eprint       = {2311.15899},
  bibsource    = {dblp computer science bibliography, https://dblp.org}
}

@article{ArmanTsaturian2017:arXiv,
      author={Andrii Arman and Sergei Tsaturian},
      title={The maximum number of cycles in a graph with fixed number of edges}, 
      journal={CoRR},
      volume={abs/1702.02662},
      year={2017},
      doi={10.48550/ARXIV.1702.02662},
      eprinttype={arXiv},
      eprint={1702.02662},
      url={https://arxiv.org/abs/1702.02662}, 
}

@article{Arcuri+2012:TSE,
  author       = {Andrea Arcuri and
                  Muhammad Zohaib Z. Iqbal and
                  Lionel C. Briand},
  title        = {Random Testing: Theoretical Results and Practical Implications},
  journal      = {{IEEE} Trans. Software Eng.},
  volume       = {38},
  number       = {2},
  pages        = {258--277},
  year         = {2012},
  url          = {https://doi.org/10.1109/TSE.2011.121},
  doi          = {10.1109/TSE.2011.121},
  bibsource    = {dblp computer science bibliography, https://dblp.org}
}

@inproceedings{Bicchierai+2023:ISSRE,
  author       = {Irene Bicchierai and Enrico Araniti and Sara Nieves Matheu{-}Garc{\'{\i}}a and Juan Francisco Mart{\'{\i}}nez Gil},
  title        = {Validating the {BIECO} Security Evaluation Methodology within a Smart Grid Monitoring {SW}},
  booktitle    = {34th International Symposium on Software Reliability Engineering, {ISSRE} - Workshops},
  address      = {Florence, Italy},
  pages        = {166--173},
  publisher    = {{IEEE}},
  year         = {2023},
  date         = {October 9-12},
  doi          = {10.1109/ISSREW60843.2023.00069},
  bibsource    = {dblp computer science bibliography, https://dblp.org}
}

@inproceedings{Castro-Cabrera+2022:ICSEA,
  title        = {A Case Study on Combining Model-based Testing and Constraint Programming for Path Coverage},
  ISBN         = {978-1-61208-997-3},
  booktitle    = {ICSEA 2022 The Seventeenth International Conference on Software Engineering Advances},
  publisher    = {ICSEA 2022 Editors Lugi Lavazza, University of Insubria at Varese, Italy},
  author       = {Maria Carmen de Castro-Cabrera and Garcia-Dominguez, Antonio and Medina-Bulo, Inmaculada},
  year         = {2022},
  pages        = {41-45},
  url          = {https://personales.upv.es/thinkmind/dl/conferences/icsea/icsea_2022/icsea_2022_1_70_10048.pdf}
}

@article{Chen+:2025:JACM,
  author       = {Li Chen and
                  Rasmus Kyng and
                  Yang P. Liu and
                  Richard Peng and
                  Maximilian Probst Gutenberg and
                  Sushant Sachdeva},
  title        = {Maximum Flow and Minimum-Cost Flow in Almost-Linear Time},
  journal      = {J. {ACM}},
  volume       = {72},
  number       = {3},
  pages        = {19:1--19:103},
  year         = {2025},
  doi          = {10.1145/3728631},
  bibsource    = {dblp computer science bibliography, https://dblp.org}
}

@inproceedings{ChetouaneWotawa2023:ICST,
  author       = {Nour Chetouane and
                  Franz Wotawa},
  title        = {Generating concrete test cases from vehicle data using models obtained
                  from clustering},
  booktitle    = {{IEEE} International Conference on Software Testing, Verification
                  and Validation, {ICST} 2023 - Workshops, Dublin, Ireland, April 16-20,
                  2023},
  pages        = {70--77},
  publisher    = {{IEEE}},
  year         = {2023},
  doi          = {10.1109/ICSTW58534.2023.00024},
  bibsource    = {dblp computer science bibliography, https://dblp.org}
}

@inproceedings{Dalal+1999:ICSE,
  author       = {Siddhartha R. Dalal and
                  Ashish Jain and
                  Nachimuthu Karunanithi and
                  J. M. Leaton and
                  Christopher M. Lott and
                  Gardner C. Patton and
                  Bruce M. Horowitz},
  editor       = {Barry W. Boehm and
                  David Garlan and
                  Jeff Kramer},
  title        = {Model-Based Testing in Practice},
  booktitle    = {Proceedings of the 1999 International Conference on Software Engineering,
                  ICSE' 99, Los Angeles, CA, USA, May 16-22, 1999},
  pages        = {285--294},
  publisher    = {{ACM}},
  year         = {1999},
  doi          = {10.1145/302405.302640},
  bibsource    = {dblp computer science bibliography, https://dblp.org}
}

@inproceedings{Darwish+2017:ICST,
  author       = {Rashid Darwish and
                  Lynnie Nakyanzi Gwosuta and
                  Richard Torkar},
  title        = {A Controlled Experiment on Coverage Maximization of Automated Model-Based
                  Software Test Cases in the Automotive Industry},
  booktitle    = {2017 {IEEE} International Conference on Software Testing, Verification
                  and Validation, {ICST} 2017, Tokyo, Japan, March 13-17, 2017},
  pages        = {546--547},
  publisher    = {{IEEE} Computer Society},
  year         = {2017},
  doi          = {10.1109/ICST.2017.65},
  bibsource    = {dblp computer science bibliography, https://dblp.org}
}

@inproceedings{DiasNeto+2007:WEASELTech,
  author = {Dias Neto, Arilo C. and Subramanyan, Rajesh and Vieira, Marlon and Travassos, Guilherme H.},
  title = {A survey on model-based testing approaches: a systematic review},
  year = {2007},
  isbn = {9781595938800},
  publisher = {Association for Computing Machinery},
  address = {New York, NY, USA},
  doi = {10.1145/1353673.1353681},
  booktitle = {Proceedings of the 1st ACM International Workshop on Empirical Assessment of Software Engineering Languages and Technologies: Held in Conjunction with the 22nd IEEE/ACM International Conference on Automated Software Engineering (ASE) 2007},
  pages = {31–36},
  numpages = {6},
  location = {Atlanta, Georgia},
  series = {WEASELTech '07}
}

@inproceedings{DwarakanathJankiti2014:ICTSS,
  author       = {Anurag Dwarakanath and
                  Aruna Jankiti},
  editor       = {Mercedes G. Merayo and
                  Edgardo Montes de Oca},
  title        = {Minimum Number of Test Paths for Prime Path and Other Structural Coverage
                  Criteria},
  booktitle    = {Testing Software and Systems - 26th {IFIP} {WG} 6.1 International
                  Conference, {ICTSS} 2014, Madrid, Spain, September 23-25, 2014. Proceedings},
  series       = {Lecture Notes in Computer Science},
  volume       = {8763},
  pages        = {63--79},
  publisher    = {Springer},
  year         = {2014},
  doi          = {10.1007/978-3-662-44857-1_5},
  bibsource    = {dblp computer science bibliography, https://dblp.org}
}

@misc{Elmendorf1970:IBM,
  title        = {Automated design of program test libraries},
  author       = {Elmendorf, William R},
  howpublished = {IBM Technial Report TR 00.2089},
  year         = {1970},
  url          = {https://www.benderrbt.com/Automated%20Design%20of%20Program%20Test%20Libraries%20-%201970.pdf}
}

@article{FordFulkerson1956:CJM,
  title        = {Maximal Flow Through a Network},
  volume       = {8},
  doi          = {10.4153/CJM-1956-045-5},
  journal      = {Canadian Journal of Mathematics},
  author       = {Ford, L. R. and Fulkerson, D. R.},
  year         = {1956},
  pages        = {399–404}
}

@inproceedings{Frey+2012:ETFA,
  author       = {Georg Frey and
                  Rainer Drath and
                  Bastian Schlich and
                  Robert Eschbach},
  title        = {"Safety automata" - {A} new specification language for the development of {PLC} safety applications},
  booktitle    = {Proceedings of 2012 17th International Conference on Emerging Technologies {\&} Factory Automation, {ETFA}},
  pages        = {1--8},
  publisher    = {{IEEE}},
  year         = {2012},
  date         = {September 17-21},
  address      = {Krakow, Poland},
  doi          = {10.1109/ETFA.2012.6489536},
  bibsource    = {dblp computer science bibliography, https://dblp.org}
}

@article{Garousi+2021:JSS,
  title = {Model-based testing in practice: An experience report from the web applications domain},
  journal = {Journal of Systems and Software},
  volume = {180},
  pages = {111032},
  year = {2021},
  issn = {0164-1212},
  doi = {https://doi.org/10.1016/j.jss.2021.111032},
  author = {Vahid Garousi and Alper Buğra Keleş and Yunus Balaman and Zeynep Özdemir Güler and Andrea Arcuri}
}

@article{Gautschi+2017:TVLSI,
  author       = {Michael Gautschi and
                  Pasquale Davide Schiavone and
                  Andreas Traber and
                  Igor Loi and
                  Antonio Pullini and
                  Davide Rossi and
                  Eric Flamand and
                  Frank K. G{\"{u}}rkaynak and
                  Luca Benini},
  title        = {Near-Threshold {RISC-V} Core With {DSP} Extensions for Scalable IoT
                  Endpoint Devices},
  journal      = {{IEEE} Trans. Very Large Scale Integr. Syst.},
  volume       = {25},
  number       = {10},
  pages        = {2700--2713},
  year         = {2017},
  doi          = {10.1109/TVLSI.2017.2654506},
  bibsource    = {dblp computer science bibliography, https://dblp.org}
}

@article{Gudmundsson+2015:ISSE,
  author       = {Gudmundsson, Vignir and Schulze, Christoph and Ganesan, Dharmalingam and Lindvall, Mikael and Wiegand, Robert},
  date         = {2015/09/01},
  doi = {10.1007/s11334-015-0254-6},
  id = {Gudmundsson2015},
  isbn = {1614-5054},
  journal = {Innovations in Systems and Software Engineering},
  number = {3},
  pages = {217--232},
  title = {Model-based testing of NASA's GMSEC, a reusable framework for ground system software},
  volume = {11},
  year = {2015}
}

@inproceedings{Gudmundsson+2016:PrePostIFM,
  author       = {Vignir Gudmundsson and
                  Mikael Lindvall and
                  Luca Aceto and
                  Johann Bergthorsson and
                  Dharmalingam Ganesan},
  editor       = {Luca Aceto and
                  Adrian Francalanza and
                  Anna Ing{\'{o}}lfsd{\'{o}}ttir},
  title        = {Model-based Testing of Mobile Systems - An Empirical Study on QuizUp
                  Android App},
  booktitle    = {Proceedings First Workshop on Pre- and Post-Deployment Verification
                  Techniques, PrePost@IFM 2016, Reykjav{\'{\i}}k, Iceland, 4th
                  June 2016},
  series       = {{EPTCS}},
  volume       = {208},
  pages        = {16--30},
  year         = {2016},
  doi          = {10.4204/EPTCS.208.2},
  bibsource    = {dblp computer science bibliography, https://dblp.org}
}

@inproceedings{Hemmati+2010:FSE,
  author       = {Hadi Hemmati and
                  Lionel C. Briand and
                  Andrea Arcuri and
                  Shaukat Ali},
  editor       = {Gruia{-}Catalin Roman and
                  Andr{\'{e}} van der Hoek},
  title        = {An enhanced test case selection approach for model-based testing:
                  an industrial case study},
  booktitle    = {Proceedings of the 18th {ACM} {SIGSOFT} International Symposium on
                  Foundations of Software Engineering, 2010, Santa Fe, NM, USA, November
                  7-11, 2010},
  pages        = {267--276},
  publisher    = {{ACM}},
  year         = {2010},
  doi          = {10.1145/1882291.1882331},
  bibsource    = {dblp computer science bibliography, https://dblp.org}
}

@inproceedings{Hufkens+2024:AST,
  author       = {Lianne V. Hufkens and Fernando Pastor Ric{\'{o}}s and Beatriz Mar{\'{\i}}n and Tanja E. J. Vos},
  editor       = {Francesca Lonetti and Antonio Guerriero and Mehrdad Saadatmand and Christof J. Budnik and Jenny Li},
  title        = {Grammar-Based Action Selection Rules for Scriptless Testing},
  booktitle    = {Proceedings of the 5th International Conference on Automation of Software Test {(AST)}},
  pages        = {56--65},
  publisher    = {{ACM}},
  year         = {2024},
  date         = {April 15-16},
  address      = {Lisbon, Portugal}, 
  doi          = {10.1145/3644032.3644446},
  bibsource    = {dblp computer science bibliography, https://dblp.org}
}

@article{Janicki+2012:STVR,
  author       = {Marek Janicki and
                  Mika Katara and
                  Tuula P{\"{a}}{\"{a}}kk{\"{o}}nen},
  title        = {Obstacles and opportunities in deploying model-based {GUI} testing
                  of mobile software: a survey},
  journal      = {Softw. Test. Verification Reliab.},
  volume       = {22},
  number       = {5},
  pages        = {313--341},
  year         = {2012},
  url          = {https://doi.org/10.1002/stvr.460},
  doi          = {10.1002/STVR.460},
  bibsource    = {dblp computer science bibliography, https://dblp.org}
}

@inproceedings{Kaminski+2010:SERP,
  author       = {Garrett Kent Kaminski and
                  Upsorn Praphamontripong and
                  Paul Ammann and
                  Jeff Offutt},
  editor       = {Hamid R. Arabnia and
                  Hassan Reza and
                  Leonidas Deligiannidis and
                  Juan Jose Cuadrado{-}Gallego and
                  Vincent Schmidt and
                  Ashu M. G. Solo},
  title        = {An Evaluation of the Minimal-MUMCUT Logic Criterion and Prime Path
                  Coverage},
  booktitle    = {Proceedings of the 2010 International Conference on Software Engineering
                  Research {\&} Practice, {SERP} 2010, July 12-15, 2010, Las Vegas,
                  Nevada, USA, 2 Volumes},
  pages        = {205--211},
  publisher    = {{CSREA} Press},
  year         = {2010},
  bibsource    = {dblp computer science bibliography, https://dblp.org},
  url          = {https://www.academia.edu/download/30711986/mumcut-prime.pdf}
}

@inproceedings{Karlsson+2021:ICSTW,
  author       = {Stefan Karlsson and
                  Adnan Causevic and
                  Daniel Sundmark and
                  M{\aa}rten Larsson},
  title        = {Model-based Automated Testing of Mobile Applications: An Industrial
                  Case Study},
  booktitle    = {14th {IEEE} International Conference on Software Testing, Verification
                  and Validation Workshops, {ICST} Workshops 2021, Porto de Galinhas,
                  Brazil, April 12-16, 2021},
  pages        = {130--137},
  publisher    = {{IEEE}},
  year         = {2021},
  doi          = {10.1109/ICSTW52544.2021.00033},
  bibsource    = {dblp computer science bibliography, https://dblp.org}
}

@inproceedings{Karlsson+2022:ACM,
  author       = {Karlsson, Viktor Aronsson and Almasri, Ahmed and Enoiu, Eduard Paul and Afzal, Wasif and Charbachi, Peter},
  title        = {Automation of the creation and execution of system level hardware-in-loop tests through model-based testing},
  year         = {2022},
  isbn         = {9781450394529},
  publisher    = {Association for Computing Machinery},
  address      = {New York, NY, USA},
  doi          = {10.1145/3548659.3561313},
  booktitle    = {Proceedings of the 13th International Workshop on Automating Test Case Design, Selection and Evaluation},
  pages        = {9–16},
  numpages     = {8},
  location     = {Singapore, Singapore},
  series       = {A-TEST 2022}
}

@article{Kilincceker+2021:IEEEAccess,
  author       = {Onur K{\i}l{\i}n{\c{c}}{\c{c}}eker and
                  Alper Silistre and
                  Fevzi Belli and
                  Moharram Challenger},
  title        = {Model-Based Ideal Testing of {GUI} Programs-Approach and Case Studies},
  journal      = {{IEEE} Access},
  volume       = {9},
  pages        = {68966--68984},
  year         = {2021},
  doi          = {10.1109/ACCESS.2021.3077518},
  bibsource    = {dblp computer science bibliography, https://dblp.org}
}

@article{Kilincceker+2022:SSM,
  author       = {Onur K{\i}l{\i}n{\c{c}}{\c{c}}eker and Ercument T{\"u}rk and Fevzi Belli and Moharram Challenger},
  title        = {Model-based ideal testing of hardware description language {(HDL)} programs},
  journal      = {Softw. Syst. Model.},
  volume       = {21},
  number       = {3},
  pages        = {1209--1240},
  year         = {2022},
  doi          = {10.1007/S10270-021-00934-6},
  bibsource    = {dblp computer science bibliography, https://dblp.org}
}

@inproceedings{Koroglu+2025:ICSTW:AMOST,
  author       = {Yavuz K{\"{o}}ro{\u{g}}lu and
                  Mutlu Beyaz{\i}t and
                  Onur K{\i}l{\i}n{\c{c}}{\c{c}}eker and
                  Serge Demeyer and
                  Franz Wo\-tawa},
  title        = {Towards Improving Automated Testing with GraphWalker},
  booktitle    = {{IEEE} International Conference on Software Testing, Verification
                  and Validation, {ICST} 2025 - Workshops, Naples, Italy, March 31 -
                  April 4, 2025},
  pages        = {54--58},
  publisher    = {{IEEE}},
  year         = {2025},
  doi          = {10.1109/ICSTW64639.2025.10962480},
  bibsource    = {dblp computer science bibliography, https://dblp.org}
}

@inproceedings{Leal+2020:PRDC,
  author       = {Lucas Leal and Leonardo Montecchi and Andrea Ceccarelli and Eliane Martins},
  title        = {Using Metamodels to Improve Model-Based Testing of Service Orchestrations},
  booktitle    = {25th Pacific Rim International Symposium on Dependable Computing, {PRDC}},
  pages        = {130--139},
  publisher    = {{IEEE}},
  year         = {2020},
  date         = {December 1-4},
  address      = {Perth, Australia},
  doi          = {10.1109/PRDC50213.2020.00024},
  bibsource    = {dblp computer science bibliography, https://dblp.org}
}

@inproceedings{Li+:2012:ICST,
  author       = {Nan Li and
                  Fei Li and
                  Jeff Offutt},
  editor       = {Giuliano Antoniol and
                  Antonia Bertolino and
                  Yvan Labiche},
  title        = {Better Algorithms to Minimize the Cost of Test Paths},
  booktitle    = {Fifth {IEEE} International Conference on Software Testing, Verification
                  and Validation, {ICST} 2012, Montreal, QC, Canada, April 17-21, 2012},
  pages        = {280--289},
  publisher    = {{IEEE} Computer Society},
  year         = {2012},
  doi          = {10.1109/ICST.2012.108},
  bibsource    = {dblp computer science bibliography, https://dblp.org}
}

@inproceedings{Lindvall+2015:ICSE,
  author       = {Mikael Lindvall and Dharmalingam Ganesan and Ragnar Ardal and Robert E. Wiegand},
  editor       = {Antonia Bertolino and Gerardo Canfora and Sebastian G. Elbaum},
  title        = {Metamorphic Model-Based Testing Applied on {NASA} {DAT} - An Experience Report},
  booktitle    = {37th International Conference on Software Engineering, {ICSE}},
  pages        = {129--138},
  publisher    = {{IEEE} Computer Society},
  year         = {2015},
  date         = {May 16-24},
  address      = {Florence, Italy}, 
  volume       = {2},
  doi          = {10.1109/ICSE.2015.348},
  bibsource    = {dblp computer science bibliography, https://dblp.org}
}

@article{MasriZaraket:2016:ElsevierAdvC,
  author       = {Wes Masri and
                  Fadi A. Zaraket},
  title        = {Coverage-Based Software Testing: Beyond Basic Test Requirements},
  journal      = {Adv. Comput.},
  volume       = {103},
  pages        = {79--142},
  year         = {2016},
  doi          = {10.1016/BS.ADCOM.2016.04.003},
  bibsource    = {dblp computer science bibliography, https://dblp.org}
}

@article{Mlynarski+2012:AdvC,
  author       = {Michael Mlynarski and
                  Baris G{\"{u}}ldali and
                  Stephan Wei{\ss}leder and
                  Gregor Engels},
  title        = {Model-Based Testing: Achievements and Future Challenges},
  journal      = {Adv. Comput.},
  volume       = {86},
  pages        = {1--39},
  year         = {2012},
  url          = {https://doi.org/10.1016/B978-0-12-396535-6.00001-6},
  doi          = {10.1016/B978-0-12-396535-6.00001-6},
  bibsource    = {dblp computer science bibliography, https://dblp.org}
}

@article{NtafosHakimi1979:TSE,
  author       = {Simeon C. Ntafos and
                  S. Louis Hakimi},
  title        = {On Path Cover Problems in Digraphs and Applications to Program Testing},
  journal      = {{IEEE} Trans. Software Eng.},
  volume       = {5},
  number       = {5},
  pages        = {520--529},
  year         = {1979},
  doi          = {10.1109/TSE.1979.234213},
  bibsource    = {dblp computer science bibliography, https://dblp.org}
}

@book{Rumpe2016:Springer,
  author       = {Bernhard Rumpe},
  title        = {Modeling with {UML}},
  publisher    = {Springer},
  year         = {2016},
  doi          = {10.1007/978-3-319-33933-7},
  isbn         = {978-3-319-33932-0},
  bibsource    = {dblp computer science bibliography, https://dblp.org}
}

@inproceedings{Schur+2013:FSE,
  author       = {Schur, Matthias and Roth, Andreas and Zeller, Andreas},
  title        = {Mining behavior models from enterprise web applications},
  year         = {2013},
  isbn         = {9781450322379},
  publisher    = {Association for Computing Machinery},
  address      = {New York, NY, USA},
  doi          = {10.1145/2491411.2491426},
  booktitle    = {Proceedings of the 2013 9th Joint Meeting on Foundations of Software Engineering},
  pages        = {422--432},
  numpages     = {11},
  location     = {Saint Petersburg, Russia},
  series       = {ESEC/FSE 2013}
}

@article{Thorup2004:JCSS:STOC,
  title        = {Integer priority queues with decrease key in constant time and the single source shortest paths problem},
  journal      = {Journal of Computer and System Sciences},
  volume       = {69},
  number       = {3},
  pages        = {330-353},
  year         = {2004},
  note         = {Special Issue on STOC 2003},
  issn         = {0022-0000},
  doi          = {10.1016/j.jcss.2004.04.003},
  author       = {Mikkel Thorup}
}

@inproceedings{Tiwari+2022:APSEC,
  author       = {Saurabh Tiwari and Kumar Iyer and Eduard Paul Enoiu},
  title        = {Combining Model-Based Testing and Automated Analysis of Behavioural Models using GraphWalker and {UPPAAL}},
  booktitle    = {29th Asia-Pacific Software Engineering Conference, {APSEC}},
  pages        = {452--456},
  publisher    = {{IEEE}},
  year         = {2022},
  date         = {December 6-9},
  address      = {Virtual Event, Japan},
  doi          = {10.1109/APSEC57359.2022.00061},
  bibsource    = {dblp computer science bibliography, https://dblp.org}
}

@article{Utting+2016:AdvC,
  author       = {Mark Utting and
                  Bruno Legeard and
                  Fabrice Bouquet and
                  Elizabeta Fourneret and
                  Fabien Peureux and
                  Alexandre Vernotte},
  title        = {Recent Advances in Model-Based Testing},
  journal      = {Adv. Comput.},
  volume       = {101},
  pages        = {53--120},
  year         = {2016},
  url          = {https://doi.org/10.1016/bs.adcom.2015.11.004},
  doi          = {10.1016/BS.ADCOM.2015.11.004},
  bibsource    = {dblp computer science bibliography, https://dblp.org}
}

@article{Williams2018:SiamComp,
  author       = {R. Ryan Williams},
  title        = {Faster All-Pairs Shortest Paths via Circuit Complexity},
  journal      = {{SIAM} J. Comput.},
  volume       = {47},
  number       = {5},
  pages        = {1965--1985},
  year         = {2018},
  doi          = {10.1137/15M1024524},
  bibsource    = {dblp computer science bibliography, https://dblp.org}
}

@inproceedings{Zafar+2023:ENASE,
  author       = {Muhammad Nouman Zafar and
                  Wasif Afzal and
                  Eduard Paul Enoiu},
  editor       = {Hermann Kaindl and
                  Mike Mannion and
                  Leszek A. Maciaszek},
  title        = {An Empirical Evaluation of System-Level Test Effectiveness for Safety-Critical
                  Software},
  booktitle    = {Proceedings of the 18th International Conference on Evaluation of
                  Novel Approaches to Software Engineering, {ENASE} 2023, Prague, Czech
                  Republic, April 24-25, 2023},
  pages        = {293--305},
  publisher    = {{SCITEPRESS}},
  year         = {2023},
  doi          = {10.5220/0011756800003464},
  bibsource    = {dblp computer science bibliography, https://dblp.org}
}

\subsubsection*{}\label{sec:bio}%
\noindent \textbf{Yavuz K{\"o}ro\u{g}lu} is an assistant professor at {\.I}stanbul Technical University. He received B.Sc., M.Sc., and Ph.D. degrees from Bo\u{g}azi\c{c}i University, T{\"u}rkiye, in 2014, 2016, and 2023, respectively, and continued his postdoctoral research studies at Graz University of Technology until 2025. His research interests include model-based testing, fuzzing, formal methods, artificial intelligence, bug prediction, and machine learning.

\subsubsection*{}%
\noindent \textbf{Mutlu Beyaz{\i}t} received a master's degree from {\.I}zmir Institute of Technology, and a Ph.D. degree from the University of Paderborn. After completing his Ph.D. study, he was employed as a full-time Faculty Member of the Department of Computer Engineering, Ya\c{s}ar University. Currently, he is working as a senior researcher at the University of Antwerp.

\subsubsection*{}%
\noindent \textbf{Onur K{\i}l{\i}n\c{c}\c{c}eker} received a master's degree from Ege University, and a Ph.D. degree from the University of Paderborn. During his master's and Ph.D. studies, he was also employed as a research assistant at the Mu\u{g}la S{\i}tk{\i} Ko\c{c}man University and Ege University and then as a researcher at the University of Antwerp. Currently, he is working as a senior researcher at the University of Antwerp. His research interests include software testing, model-based testing, and mutation testing.

\subsubsection*{}%
\noindent \textbf{Serge Demeyer}
is a professor at the University of Antwerp and the spokesperson for the NEXOR research consortium on cyber-physical systems. He directs a research lab investigating the theme of ``Software Reengineering'' (LORE - Lab On REengineering). Serge Demeyer received a ``Best Teachers Award" from the Faculty of Sciences at the University of Antwerp and is still very active in all matters related to teaching quality. His main research interest concerns software test automation and how this helps in striking the right balance between reliability (striving for perfection) and agility (optimising for adaptability). He is an active member of the corresponding international research communities, serving in various conference organization and program committees. He has written a book entitled ``Object-Oriented Reengineering'' and edited a book on ``Software Evolution''. He also authored numerous peer reviewed articles, many of them in top conferences and journals.

\subsubsection*{}%
\noindent \textbf{Franz Wotawa} received an M.Sc. in Computer Science (1994) and a Ph.D. in 1996, both from the Vienna University of Technology. He is currently a professor of software engineering at the Graz University of Technology and the head of the Institute of Software Engineering and Artifical Intelligence. His research interests include model-based and qualitative reasoning, theorem proving, mobile robots, verification and validation, and software testing and debugging.

\end{document}